\documentstyle[12pt,epsf,a4]{article}
\newcommand{\gsim}{\;\rlap{\lower 3.5 pt \hbox{$\mathchar \sim$}} \raise 1pt
 \hbox {$>$}\;}
\newcommand{\lsim}{\;\rlap{\lower 3.5 pt \hbox{$\mathchar \sim$}} \raise 1pt
 \hbox {$<$}\;}

\renewcommand{\thefootnote}{\fnsymbol{footnote}}
\newcommand{\qsla}{q\hspace{-.5em/\hspace{.5em}}}
\newcommand{\lnmum}{l_{\mu M}}
\newcommand{\lmM}{l_{\mu M}}
\newcommand{\lmm}{l_{\mu m}}
\newcommand{\logfu}{l_{\mu_f}}
\newcommand{\lmucft}{l_{\mu,t}}

\begin{document}    

\title{\vskip-3cm{\baselineskip14pt
\centerline{\normalsize\hfill DESY 99--174}
\centerline{\normalsize\hfill TTP99--47}
\centerline{\normalsize\hfill hep-ph/9911434}
\centerline{\normalsize\hfill November 1999}
}
\vskip.7cm
The relation between the $\overline{\rm MS}$ and the on-shell quark mass
at order $\alpha_s^3$
\vskip.3cm
}
\author{
{K.G. Chetyrkin}$^{a,}$\thanks{Permanent address:
Institute for Nuclear Research, Russian Academy of Sciences,
60th October Anniversary Prospect 7a, Moscow 117312, Russia.}
\,and
{M. Steinhauser}$^b$
  \\[3em]
 { (a) Institut f\"ur Theoretische Teilchenphysik,}\\
  { Universit\"at Karlsruhe, D-76128 Karlsruhe, Germany}
  \\[.5em]
  { (b) II. Institut f\"ur Theoretische Physik,}\\ 
  { Universit\"at Hamburg, D-22761 Hamburg, Germany}
}
\date{}
\maketitle

\begin{abstract}
\noindent
The relation between the on-shell and $\overline{\rm MS}$ mass can be 
expressed through scalar and vector part of the quark propagator.  In
principle these two-point functions have to be evaluated on-shell
which is a non-trivial task at three-loop order.  Instead, we evaluate
the quark self energy in the limit of large and small external
momentum and use conformal mapping in combination with Pad\'e
improvement in order to construct a numerical approximation for the
relation~\cite{CheSte99}.
The errors of our final result are conservatively estimated
to be below 3\%. The numerical implications of the results are
discussed in particular in view of top and bottom quark production
near threshold.  We show that the knowledge of new ${\cal
O}(\alpha_s^3)$ correction leads to a significant reduction of the
theoretical uncertainty in the determination of the quark masses.
\end{abstract}

\thispagestyle{empty}
\newpage
\setcounter{page}{1}


\renewcommand{\thefootnote}{\arabic{footnote}}
\setcounter{footnote}{0}

\section{Introduction}

In higher order calculations there is in general an ambiguity in the prediction
for the physical quantities which can be traced back to the adopted
renormalization scheme. Different renormalization conditions imply different
numerical values for the parameters of the underlying theory.
Very often it is useful to convert them from one scheme into an other in order
to compare the final predictions for the observables in both schemes.

In Quantum Chromodynamics (QCD)
practical calculations are very often performed in the
modified minimal subtraction ($\overline{\rm MS}$)
scheme~\cite{tHo73,BarBurDukMut78}
leading to the definition of the so-called short-distance $\overline{\rm MS}$
mass.
The $\overline{\rm MS}$ mass occupies a distinguished place among
various mass definitions. First, it is a truly short distance mass not
suffering from nonperturbative ambiguities. Second, the $\overline{\rm
MS}$ mass proves to be extremely convenient in multi-loop calculations
of mass-dependent inclusive physical observables dominated by short
distances (for a review see~\cite{ckk96}). 
On the other hand the experiments often provide masses
which are tightly connected to the on-shell definition.
Thus, conversion formulae
are needed in order to make contact between theory and experiment.
The two-loop relation between $\overline{\rm MS}$ and the on-shell definition
of the quark mass has been obtained in~\cite{GraBroGraSch90}
and has been confirmed in~\cite{FleJegTarVer99}.
Until recently the accuracy of this equation was enough for the practical
applications.
Meanwhile, however, new computations have become available which require the
relation between the $\overline{\rm MS}$ and on-shell mass at ${\cal
  O}(\alpha_s^3)$ in order to perform a consistent analysis. 
The necessity of an accurate determination of the quark masses,
especially those of the top and bottom ones,
is demonstrated by the following two examples.

The main goal of the future $B$ physics experiments is the
determination of the Cabibbo-Kobayashi-Maskawa matrix elements which will
give deeper insight into the origin of CP violation and possibly also 
provides hints to new physics.
In particular the precise measurement of $V_{cb}$ is very promising.
It is determined from semileptonic $B$ meson
decay rates. Thus it is desirable to know the bottom quark mass
as accurately as possible as it enters already the Born result to the fifth
power.

One of the primary goals of a future electron-positron linear collider
or muon collider will be the precise determination of the top quark
properties, especially its mass, $M_t$.
In hadron colliders like the Fermilab TEVATRON or
the Large Hadron Collider the top quarks are reconstructed from
the invariant mass of the $W$ bosons and the bottom quarks.
On the contrary in lepton colliders
it is possible to determine the top quark mass from the line shape of the
production cross section $\sigma(e^+ e^- \to t\bar{t})$ close to the
threshold.
Simulation studies have shown that an experimental uncertainty
of about $100$~MeV in the top mass determination can be
achieved~\cite{sim95}.
Thus also from the theoretical side the ambiguities have to be
controlled with the same precision.

Threshold phenomena are conveniently expressed in terms of 
the pole mass, $M$, of the corresponding quark
as this parameter naturally appears in the equations which have to be solved.
The pole mass is a gauge invariant, infrared-finite and a
renormalization scheme independent
quantity~\cite{Tar81,Nar87,Bre95,Kro98}.
However, various calculations~\cite{ttNNLO} have shown that
higher order corrections have an significant influence on the 
predictions for the quark production cross section.
It has been shown that this is connected to the definition of the
quark mass used for the parameterization of the cross
section~\cite{Ben98,HoaSmiSteWil98}.
A common remedy is the introduction of a new mass parameter
which is supposed to be insensitive to long distance phenomena.
One thus often
speaks about short-distance masses. The relation of the new mass parameter
to the pole mass is used in order to re-parameterize the threshold phenomena.
On the other hand a relation of the new quark mass to the $\overline{\rm MS}$
mass must be established as
it is commonly used for the parameterization of those quantities which are not
related to the threshold.

Recently several groups evaluated next-to-next-to-leading order
results for the cross section $\sigma(e^+e^-\to t\bar{t})$
close to threshold~\cite{ttNNLO,BenSigSmi99,NagOtaSum99,PenPiv99,HoaTeu99}.
In order to establish a consistent
relation to the $\overline{\rm MS}$ mass the three-loop relation between the
$\overline{\rm MS}$ and the on-shell mass is needed. This will be
discussed in Section~\ref{sec:appl}.

The bottom quark mass can be obtained from sum rules for the
$\Upsilon$ mesons~\cite{Vol1,Vol2}. The current strategy can be
schematically outlined as follows~\cite{Vol,JP,Kuehn:npb534}.  Moments
of the current correlator involving bottom quarks are determined with
the help of experimental results on the electronic decay width and the
masses of the $\Upsilon$ mesons.  They are compared with the moments
computed with the help of non-relativistic QCD. Due to the
strong dependence on the bottom quark mass a precise determination is
possible.  Recently a next-to-next-to-leading order computation has
become available~\cite{PenPiv98,Hoa99}.  Again, for a consistent
determination of the $\overline{\rm MS}$ mass the order $\alpha_s^3$
relation to the on-shell mass is necessary.

A somewhat different approach for the determination of both the charm and
bottom mass has been performed in~\cite{PinYnd98}.
The lower states in the heavy quarkonium spectrum have been computed
up to order $\alpha_s^4$ which again demands for the three-loop
$\overline{\rm MS}$--on-shell mass relation in order to
evaluate the $\overline{\rm MS}$ mass.

Another application where the three-loop relation between the on-shell and the
$\overline{\rm MS}$ mass is needed is connected the recent evaluation of
the quartic mass corrections at order $\alpha_s^3$ to the
cross section $\sigma(e^+e^-\to\mbox{hadrons})$~\cite{Har:diss,CheHarKue99}.
For conceptual reasons the calculation has been performed using the
$\overline{\rm MS}$ mass. If, however, the result should be expressed in terms
of the pole mass the corresponding relation has to be known to order
$\alpha_s^3$. Note the Born result expanded for small quark masses doesn't
have a quadratic term. Thus in this case
the mass relation is only needed up to order $\alpha_s^2$
and the quartic corrections require for the first time the ${\cal
  O}(\alpha_s^3)$ terms.

In the present article we describe the calculation of 
the ${\cal O}(\alpha_s^3)$ correction to the  mass 
relation as well as the implications of the result. A short version
containing the main results has been published in~\cite{CheSte99}. 

The outline of the paper is as follows. In the next section the notation is
introduced. Afterwards in Section~\ref{sec:method} the method we use for the
calculation is described in detail. Its power is demonstrated in
Section~\ref{sec:12loop} at the one- and two-loop level where a comparison
with the exact results is possible.
The tools needed for the three-loop calculation are presented in
Section~\ref{sec:master}. In particular we collect all
difficult integrals which are needed for the computation of the fermion
propagator. The three-loop results for the mass relation are finally discussed
in Section~\ref{sec:3loop} and their implications to quark production
processes are mentioned in Section~\ref{sec:appl}.
Our conclusions are presented in Section~\ref{sec:con}.


\section{Notation}

This section is devoted to set up the notation.
Throughout this paper the bare (unrenormalized) mass is denoted by
$m^0$ and the $\overline{\rm MS}$ and on-shell
renormalized ones by $m$ and $M$, respectively.
Their connection is given by the following equations
\begin{eqnarray}
m(\mu) &=& Z_m m^0 \,\,=\,\,z_m(\mu) M
\,,
\label{eq:mmsos}
\end{eqnarray}
where $z_m$ is finite and has an explicit dependence on the
renormalization 
scale $\mu$. The main goal of this paper
is the computation of $z_m$ up to order $\alpha_s^3$.
Therefore three-loop corrections to the fermion propagator have to be
considered. A convenient variable in this context is
\begin{eqnarray}
z&=&\frac{q^2}{M^2}
\,,
\end{eqnarray}
where $q$ is the external momentum of the fermion propagator.

The inverse quark propagator is denoted by
\begin{eqnarray}
  \left( S_F^0(q) \right)^{-1} &=& i \left[ m^0 \left( 1 - \Sigma_S^0 \right) 
  - \qsla \left( 1 + \Sigma_V^0 \right)\right]
\label{eq:sfinv0}
\,,
\end{eqnarray}
where the functions $\Sigma_S^0$ and $\Sigma_V^0$
depend on the external momentum $q$, the bare mass $m^0$ and on the bare
strong coupling constant $\alpha_s^0$.
The renormalized version can be cast in the form
\begin{eqnarray}
  \left(S_F(q)\right)^{-1} &=& i \left[
    \left(M-\qsla\right) S_V(z) + M \left(z_m(\mu) S_S(z) -S_V(z) \right)
  \right]
    \,,
\label{eq:sfinv}
\end{eqnarray}
with\footnote{Note that in contrast to the quantities defined
  in~\cite{CheHarSeiSte99}
  the wave function renormalization for functions $S_{S/V}$
  is still defined in the $\overline{\rm MS}$ scheme.}
\begin{eqnarray}
  S_V(z) &=& Z_2(1+\Sigma^0_V)
  \,,
  \nonumber\\
  S_S(z) &=& Z_2Z_m(1-\Sigma^0_S)
  \,.
  \label{eq:SvSs}
\end{eqnarray}
$Z_2$ denotes the wave function renormalization in the $\overline{\rm MS}$
scheme which is sufficient for our considerations.
Note that the functions $S_S$ and $S_V$ are $\overline{\rm MS}$ quantities
which later on are expressed in terms of the on-shell mass.
The two-loop relation between $m$ and $M$ is enough to
do this at order $\alpha_s^3$.

It is convenient to write the functions $S_{S/V}$ in the following way:
\begin{eqnarray}
S_{S/V} = 1 
+ \sum_{n\ge1} S_{S/V}^{(n)} \left(\frac{\alpha_s}{\pi}\right)^n
\label{eq:decas}
\,,
\end{eqnarray}
where the quantities $S_{S/V}^{(n)}$ exhibit the following colour
structures
(the indices $S$ and $V$ are omitted in the following):
\begin{eqnarray}
S^{(1)} &=& C_F S_F 
\,,
\nonumber\\
S^{(2)} &=& C_F^2 S_{FF} + C_FC_A S_{FA} + C_FTn_l S_{FL} + C_FT S_{FH}
\,,
\nonumber\\
S^{(3)} &=& C_F^3 S_{FFF} + C_F^2C_A S_{FFA} + C_FC_A^2 S_{FAA} 
          + C_F^2Tn_l S_{FFL}   + C_F^2T S_{FFH} 
\nonumber\\&&\mbox{}
          + C_FC_ATn_l S_{FAL}  + C_FC_AT S_{FAH}
          + C_FT^2n_l^2 S_{FLL} + C_FT^2n_l S_{FLH} 
\nonumber\\&&\mbox{}
          + C_FT^2 S_{FHH}
\,.
\label{eq:deccf}
\end{eqnarray}
The same decomposition also holds for the function $z_m$.
In~(\ref{eq:deccf}) $n_l$ represents the number of light (massless) quark
flavours. $C_F$ and $C_A$ are the Casimir operators of the fundamental and
adjoint representation. In the case of $SU(N_c)$ they are given by
$C_F=(N_c^2-1)/(2N_c)$ and $C_A=N_c$. The trace normalization of the
fundamental representation is $T=1/2$.
The subscripts $F$, $A$ and $L$ in Eq.~(\ref{eq:deccf})
shall remind on the colour factors $C_F$,
$C_A$ and $Tn_l$, respectively. $H$ simply stands for the
colour factor $T$.

At one- and two-loop order the results for $z_m$
read~\cite{Tar81,GraBroGraSch90}
\begin{eqnarray}
z_{m}^F(\mu) &=& -1 - \frac{3}{4}\lnmum
\,,
\nonumber\\
z_{m}^{FF}(\mu) &=&  \frac{7}{128} - \frac{15}{8}\zeta_2 
- \frac{3}{4}\zeta_3 + 3\zeta_2\ln2
+\frac{21}{32}\,\lnmum
+\frac{9}{32}\,\lnmum^2
\,\,\approx\,\,-0.51056
\,,
\nonumber\\
z_{m}^{FA}(\mu) &=& -\frac{1111}{384} + \frac{1}{2}\zeta_2 
+ \frac{3}{8}\zeta_3
- \frac{3}{2}\zeta_2\ln2  
-\frac{185}{96}\,\lnmum
-\frac{11}{32}\,\lnmum^2
\,\,\approx\,\,-3.33026
\,,
\nonumber\\
z_{m}^{FL}(\mu) &=& \frac{71}{96} + \frac{1}{2}\zeta_2
+\frac{13}{24}\,\lnmum
+\frac{1}{8}\,\lnmum^2
\,\,\approx\,\,1.56205
\,,
\nonumber\\
z_{m}^{FH}(\mu) &=& \frac{143}{96} - \zeta_2
+\frac{13}{24}\,\lnmum
+\frac{1}{8}\,\lnmum^2
\,\,\approx\,\,-0.15535
\,,
\label{eq:zm12loop}
\end{eqnarray}
with $\lnmum=\ln \mu^2/M^2$
where after the approximation signs
the choice $\mu^2=M^2$ has been adopted.
For later use we also define the propagator of the gluon.
In 't~Hooft-Feynman gauge it is given by
\begin{eqnarray}
  D_g(q) &=&
  i\frac{-g^{\mu\nu}+\xi\frac{q^\mu q^\nu}{q^2}}{q^2+i\epsilon}
  \,.
  \label{eq:gluprop}
\end{eqnarray}


\section{Method}
\label{sec:method}

A formula which allows for the computation 
of the $\overline{\rm MS}$--on-shell relation for the quark
mass is obtained from the requirement that the inverse fermion propagator has
a zero at the position of the on-shell mass:
\begin{eqnarray}
\left(S_F(q)\right)^{-1}\bigg|_{q^2=M^2} &=& 0
\,.
\label{eq:oscond}
\end{eqnarray}
At order $\alpha_s^3$ this requires the evaluation of three-loop
on-shell integrals. Currently it is quite unhandy to deal with such kind of
diagrams as up to now the literature lacks of a useful description
of an algorithm for their computation.
Even though the technology is in principle available as it was
needed for the computation of the anomalous magnetic
moment of the muon~\cite{LapRem96}.

We have decided to choose a different way for the practical calculation.
The starting point is, of course, also Eq.~(\ref{eq:oscond}).
Applying it to Eq.~(\ref{eq:sfinv}) leads to the
condition
\begin{eqnarray}
f(z) &\equiv& z_m(\mu)\,S_S(z) - S_V(z) \,\,=\,\, 0 \qquad \mbox{for}
\qquad z=1
\,.
\label{eq:f}
\end{eqnarray}
At a given loop-order $L$ the Eqs.~(\ref{eq:SvSs})
are inserted and the resulting equation is solved for $z_m^{(L)}$.
Thus Eq.~(\ref{eq:f}) can be cast in the form
\begin{eqnarray}
f(z) &=& g(z) + z_m^{(L)} \left(\frac{\alpha_s}{\pi}\right)^L
\,.
\end{eqnarray}
Our aim is the computation of $g(1)$.
Note that the individual self energies $\Sigma_S$ and $\Sigma_V$
develop infra-red singularities when they are evaluated on-shell.
The proper combination which leads to the relation between the 
$\overline{\rm MS}$ and on-shell mass is, however, free of infra-red
problems.

At this point a comment in connection to Eq.~(\ref{eq:f}) is in order.
In fact also more involved equations could be chosen which
result in more complicated expressions for $g(z)$. To us the choice
in~(\ref{eq:f}) appears very natural. Furthermore the one- and two-loop
results are reproduced with rather high accuracy.
We also tried various other options; the final
results, however, remained the same.

The strategy for the computation of $g(1)$ is as follows~\cite{pade}:
Expansions for small and large external
momentum are computed for the quark self energies.
After building the proper combinations needed for $g(z)$ a conformal
mapping~\cite{FleTar94}
is performed
\begin{eqnarray}
  z &=& \frac{4\omega}{(1+\omega)^2}
  \,,
  \label{eq:confmap} 
\end{eqnarray}
which maps the complex $z$-plane into the interior of the unit circle in
the $\omega$-plane. The relevant point $z = 1$ is mapped to $\omega = 1$.
The motivation for this conformal mapping is based on
the observation that the application
of a Pad\'e approximation relies heavily on analytic properties.
Actually, $g(z)$
develops a branch cut along the real $z$-axis starting from $z=1$. This
cut is mapped through Eq.~(\ref{eq:confmap}) onto the unit circle
of the $\omega$ plane. Thus by applying Eq.~(\ref{eq:confmap}) the radius of
convergence is enlarged.
In order to ensure also the convergence at the boundary a Pad\'e approximation
is performed. In the variable $\omega$ it is defined through
\begin{eqnarray}
[m/n](\omega) &=& \frac{a_0+a_1\omega+\ldots+a_m\omega^m}
                       {1  +b_1\omega+\ldots+b_n\omega^n}
\,.
\label{eq:padedef}
\end{eqnarray}
The details for the construction of the Pad\'e approximants can be found
in~\cite{CheKueSte96}.
Special care has to be taken with the logarithms $\ln(-q^2)$ which occur
in the results of the high-energy expansions as they could destroy the
convergence properties of the
described procedure. Details on their treatment can be found
in~\cite{CheKueSte96,CheHarSte98}.

The described procedure is applied to each colour structure occurring
in $g(z)$ separately. If we assume that the decomposition introduced in
Eqs.~(\ref{eq:decas}) and~(\ref{eq:deccf})
also holds for $g(z)$ the following equations are obtained
\begin{eqnarray}
&& g_F(z) \,\,=\,\, S_{S,F}(z) - S_{V,F}(z) \,,
\qquad
\qquad
\quad
\,\,
g_{FA}(z) \,\,=\,\, S_{S,FA}(z) - S_{V,FA}(z) \,,
\nonumber\\
&& g_{FL}(z) \,\,=\,\, S_{S,FL}(z) - S_{V,FL}(z) \,,
\qquad
\qquad
g_{FH}(z) \,\,=\,\, S_{S,FH}(z) - S_{V,FH}(z) \,,
\nonumber\\
&& g_{FF}(z) \,\,=\,\, S_{S,FF}(z) - S_{V,FF}(z) + z_m^F S_{S,F} \,,
\nonumber\\
&& g_{FAA}(z) \,\,=\,\, S_{S,FAA}(z) - S_{V,FAA}(z) \,,
\qquad
\,\,
g_{FAL}(z) \,\,=\,\, S_{S,FAL}(z) - S_{V,FAL}(z) \,,
\nonumber\\
&& g_{FAH}(z) \,\,=\,\, S_{S,FAH}(z) - S_{V,FAH}(z) \,,
\qquad
g_{FLL}(z) \,\,=\,\, S_{S,FLL}(z) - S_{V,FLL}(z) \,,
\nonumber\\
&& g_{FLH}(z) \,\,=\,\, S_{S,FLH}(z) - S_{V,FLH}(z) \,,
\qquad
g_{FHH}(z) \,\,=\,\, S_{S,FHH}(z) - S_{V,FHH}(z) \,,
\nonumber\\
&& g_{FFF}(z) \,\,=\,\, S_{S,FFF}(z) - S_{V,FFF}(z)
+ z_m^F S_{S,FF} + z_m^{FF}S_{S,F}
\,,
\nonumber\\
&& g_{FFA}(z) \,\,=\,\, S_{S,FFA}(z) - S_{V,FFA}(z)
+ z_m^F S_{S,FA} + z_m^{FA}S_{S,F} \,,
\nonumber\\
&& g_{FFL}(z) \,\,=\,\, S_{S,FFL}(z) - S_{V,FFL}(z)
+ z_m^F S_{S,FL} + z_m^{FL}S_{S,F}
\,,
\nonumber\\
&& g_{FFH}(z) \,\,=\,\, S_{S,FFH}(z) - S_{V,FFH}(z)
+ z_m^F S_{S,FH} + z_m^{FH}S_{S,F}
\,.
\label{eq:gcf}
\end{eqnarray}
The colour structures $FF$, $FFF$, $FFA$, $FFL$ and $FFH$
contain next to linear terms also products of lower order contributions.

It was already realized in~\cite{CheKueSte96} that the Pad\'e procedure
described above shows less stability as soon as diagrams are involved which
exhibit more than one particle threshold. In our case the interest is
in the lowest particle cut which happens to be for $q^2=M^2$.
The Pad\'e method heavily relies on the combination of expansions in the small
and large momentum region. The large momentum expansion, however, is
essentially sensitive to the highest particle threshold. Thus, if this
threshold numerically dominates the lower-lying ones it
cannot be expected that the Pad\'e approximation leads to stable results.
In such cases a promising alternative to the above method is the one where
only the expansion terms for $q^2\to0$ are taken into account in order to
obtain a numerical value at $q^2=M^2$.
This significantly reduces the calculational effort as
the construction of the Pad\'e approximation from low-energy moments alone is
much simpler.
In practice this approach will be applied if the Pad\'e results involving
also the high-energy data looks ill-behaved.

\begin{figure}[t]
  \begin{center}
    \begin{tabular}{c}
      \leavevmode
      \epsfxsize=14.cm
      \epsffile[70 530 530 740]{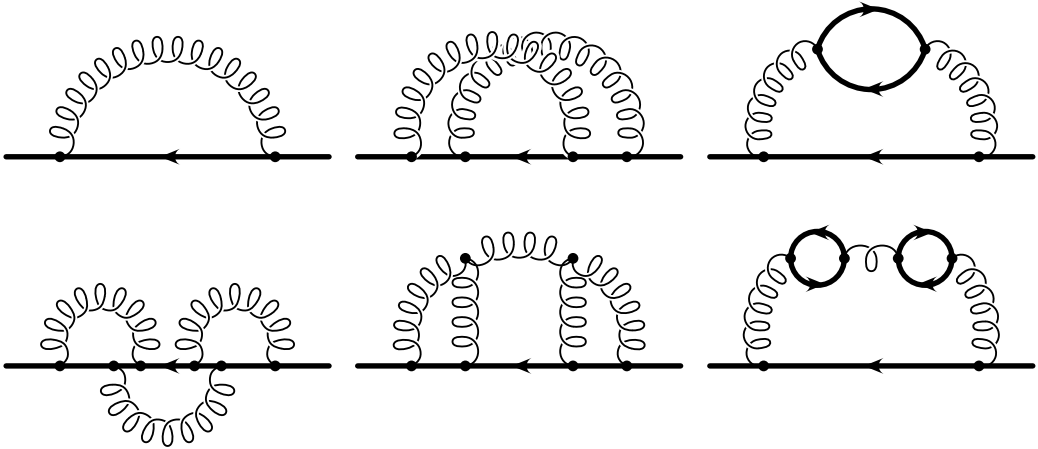}
    \end{tabular}
  \end{center}
  \caption{\label{fig:diags}
    Sample diagrams contributing to the fermion propagator.
    The fermions in the closed loops may either also carry mass $M$ or
    considered to be massless.
    }
\end{figure}

In the present analysis diagrams with other cuts than for $q^2=M^2$
are already present at the two-loop level (see Fig.~\ref{fig:diags})
which allows
us to test these suggestions.
Also at three-loop order either $q^2=M^2$ or $q^2=9M^2$ cuts
appear. Cuts involving five or more fermion lines are first possible starting
from four-loop order. Note that cuts involving an even number of fermions
cannot occur.

Let us summarize the
strategy for the practical computation: The scalar and vector
part of the one-, two- and three-loop fermion propagator is
computed. Afterwards we construct with the help of the $\overline{\rm MS}$
renormalization constants $Z_2$ and $Z_m$ and the
two-loop relation between the $\overline{\rm MS}$ and the on-shell mass
the finite functions
$S_S(z)$ and $S_V(z)$.
The proper combination according to Eqs.~(\ref{eq:gcf})
serves as a convenient starting point
to apply the conformal mapping~(\ref{eq:confmap}) and subsequently the Pad\'e
approximation in order to obtain numerical results for the
coefficients $z_m^X$ at the scale $\mu^2=M^2$. The use of the renormalization
group equation leads to expression for different choices of $\mu^2$.

As our approach relies on approximations some words concerning the error
estimate are in order.
Even at three-loop order seven terms both in the small- and large-momentum
expansion could be evaluated. Thus we restrict ourselves to those
Pad\'e results which include at least terms of order $z^5$ and $1/z^4$.
Furthermore we require that the difference of the polynomial in the numerator
and denominator is not too large, i.e. we select those Pad\'e results which
are close to the diagonal ones.
From the results we compute the average and from the spread the error
is estimated. The details will be specified below.
In the next Section we will demonstrate that this
prescription works very well at one- and two-loop order.


\section{Considerations at one- and two-loop level}
\label{sec:12loop}

Let us at this point discuss the one- and two-loop results in great detail
in order to get some feeling about the numerical quality of our procedure.
In a first step we consider Feynman gauge and demonstrate afterwards how the
results can be improved by considering a general gauge parameter.

In Tab.~\ref{tab:12loop} the results for different Pad\'e approximations
are listed. $n$ indicates the number of low-energy moments involved in the
analysis, i.e. $n=6$ implies the inclusion of terms of ${\cal O}(z^6)$.
The number of high-energy terms can be obtained in combination with
the order of the Pad\'e approximant ($[x/y]$) and is given by $x+y+1-n$.
{\footnotesize
\begin{table}[t]
\begin{center}
\begin{tabular}{|l|l||r|r|r|r|r|} 
\hline
$n$ & P.A.& $ F $& $ FF $& $ FA $& $ FL $& $ FH $\\
\hline
$5$ & $[4/5]$ &  ${}^\star-1.0010$ &  $-0.4881$ &  $-3.3813$ &  ${}^\star1.5687$ &  $-0.1613$ \\
$5$ & $[4/6]$ &  $-1.0006$ &  $-0.4862$ &  ${}^\star-3.4099$ &  $1.5666$ &  $-0.0811$ \\
$5$ & $[5/4]$ &  $-1.0008$ &  $-0.4870$ &  $-3.3936$ &  $1.5676$ &  $-0.1600$ \\
$5$ & $[5/5]$ &  $-1.0006$ &  $-0.4862$ &  $-3.3699$ &  $1.5663$ &  ${}^\star-0.1616$ \\
$5$ & $[5/6]$ &  $-1.0002$ &  ${}^\star-0.4884$ & --- &  $1.5638$ &  ${}^\star-0.1534$ \\
$5$ & $[6/4]$ &  $-1.0006$ &  $-0.4862$ &  $-3.3884$ &  $1.5665$ &  $-0.1381$ \\
$5$ & $[6/5]$ & --- &  ${}^\star-0.4884$ &  ${}^\star-3.3972$ & --- &  ${}^\star-0.1526$ \\
\hline 
$6$ & $[4/6]$ &  $-1.0007$ &  ${}^\star-0.4888$ &  $-3.3148$ &  $1.5668$ &  ${}^\star-0.1608$ \\
$6$ & $[5/5]$ &  $-1.0006$ &  $-0.4910$ &  $-3.3628$ &  $1.5665$ &  ${}^\star-0.1561$ \\
$6$ & $[5/6]$ &  $-1.0002$ &  ${}^\star-0.4866$ &  ${}^\star-3.4084$ &  $1.5640$ &  ${}^\star-0.1545$ \\
$6$ & $[5/7]$ &  $-1.0003$ &  $-0.4938$ &  ${}^\star-3.3824$ &  $1.5646$ &  ${}^\star-0.1572$ \\
$6$ & $[6/4]$ &  $-1.0007$ &  $-0.4894$ &  ${}^\star-3.3830$ &  $1.5667$ &  $-0.1382$ \\
$6$ & $[6/5]$ & --- &  ${}^\star-0.4865$ &  ${}^\star-3.4208$ & --- &  ${}^\star-0.1543$ \\
$6$ & $[6/6]$ &  $-1.0003$ &  ${}^\star-0.4897$ & --- &  $1.5646$ & --- \\
$6$ & $[7/5]$ &  $-1.0004$ & --- &  $-3.3777$ &  $1.5648$ &  ${}^\star-0.1576$ \\
\hline 
\end{tabular}
\caption{\label{tab:12loop}
  Results for the one- and two-loop coefficients of $z_m$. $n$ indicates the
  number of moments included into the analysis. Pad\'e approximations which
  develop a pole for $|\omega|<1$ are represented by a dash. Those
  where a cancellation with the numerator takes place (see text) are marked by
  a star ($\star$). For the gauge
  parameter $\xi=0$ has been adopted.
}
\end{center}
\end{table}
}

Some Pad\'e approximants develop poles inside the unit circle
($|\omega|\le1$). In general we will
discard such results in the following and represent the results by a dash.
In some cases, however, the
pole coincides with a zero of the numerator up to several digits
accuracy. These Pad\'e approximations will be marked by a star ($\star$) 
and will be taken into account in constructing our results.
To be precise: in addition
to the Pad\'e results without any poles inside the unit circle, we
will use the ones where the poles are accompanied by zeros within a
circle of radius 0.01, and the distance between the pole and the
physically relevant point $q^2/M^2=1$ is larger than 0.1.

The exact values are given in Eqs.~(\ref{eq:zm12loop}).
At first sight good agreement is obtained in all five cases.
A closer look shows that at order $\alpha_s$ almost
all results are consistently 
below the exact value. However, the maximal deviation is far below the
per mille level.
Similar observations can be made in the case of the light-fermion corrections
at order $\alpha_s^2$ which are proportional to $C_FTn_l$.
Here the deviation from the exact result amount to a few per mille.
The spread in the $C_F^2$ term at two-loop order amounts to about 1\%,
however, the exact result is not covered. The deviation is small and well
below 2\%.
For the structure $C_FC_A$ the error is smaller and the correct result
lies inside the interval.

The situation is less pleasant for the $FH$ part.
Strong fluctuations are observed in Tab.~\ref{tab:12loop}.
It seems that there is a significant influence from the
threshold at $q^2=9M^2$ which has the consequence that no stabilization is
observed if higher order terms in the large-$z$ expansion are incorporated
into the Pad\'e analysis.
Under these circumstances it is very promising,
as was already mentioned above, to ignore the high-energy terms completely and
compute the Pad\'e approximations including only the knowledge from $z\to0$.
In this case a conformal mapping is not mandatory and 
the computation can also directly be performed in the variable $z$.
The results can be found in Tab.~\ref{tab:FHzw} where the Pad\'e results which
develop poles inside the unit circle are again represented by a dash.
{\footnotesize
\begin{table}[t]
\begin{center}
\begin{tabular}{|l|l||r|r|} 
\hline
$n$ & P.A.& $ FH-z $& $ FH-\omega $\\
\hline
$4$ & $[1/3]$ &  $-0.1552$ &  $-0.1556$ \\
$4$ & $[2/2]$ &  $-0.1553$ &  $-0.1551$ \\
$4$ & $[3/1]$ &  $-0.1553$ & --- \\
$4$ & $[4/0]$ &  $-0.1552$ &  $0.0114$ \\
\hline 
$5$ & $[1/4]$ &  $-0.1553$ &  $-0.1552$ \\
$5$ & $[2/3]$ &  $-0.1553$ &  $-0.1553$ \\
$5$ & $[3/2]$ &  $-0.1553$ &  $-0.1553$ \\
$5$ & $[4/1]$ &  $-0.1553$ & --- \\
$5$ & $[5/0]$ &  $-0.1553$ &  $-0.3222$ \\
\hline 
$6$ & $[1/5]$ &  $-0.1553$ &  $-0.1554$ \\
$6$ & $[2/4]$ &  $-0.1553$ &  $-0.1554$ \\
$6$ & $[3/3]$ &  $-0.1553$ &  $-0.1555$ \\
$6$ & $[4/2]$ &  $-0.1553$ &  $-0.1554$ \\
$6$ & $[5/1]$ &  $-0.1553$ &  $-0.1568$ \\
$6$ & $[6/0]$ &  $-0.1553$ &  $0.0061$ \\
\hline 
\end{tabular}
\caption{\label{tab:FHzw}
  Results for the structure $FH$ where only low-energy moments have
  been incorporated. Only in the case of ``$FH-\omega$'' a conformal mapping
  has been used.
}
\end{center}
\end{table}
}
Note that all possible Pad\'e approximants are included --- even the 
simple Taylor expansions which correspond to the Pad\'es with 
degree zero in the denominator.
Both the stability and the agreement with the exact result is very 
impressive in the case where no conformal mapping is performed.
In the case with conformal mapping
the Taylor expansions show large oscillations.
However,
the other results nicely group around the exact value with an uncertainty
in the per mille level.
For the choice $\mu^2=M^2$ the threshold at $q^2=M^2$ is very much suppressed
as compared to the one at $q^2=9M^2$. 
This is the reason that $z$ is a very
good expansion parameter for $z\to0$ and
the sole inclusion of the low-energy terms leads
to an excellent agreement with the exact value.
Thus for $FH$ we discard the results presented in Tab.~\ref{tab:12loop}
and take the ones of Tab.~\ref{tab:FHzw} where the Pad\'e approximation was
performed in the variable $z$.

At this point some comments in connection with the Pad\'e method are in
order.
The final values for the mean and the errors actually depend on the
Pad\'e approximants included into the analysis. In Tab.~\ref{tab:12loop}
only those results are incorporated which involve $z^5$,
respectively, $z^6$ terms and corrections of order
$1/z^4$, $1/z^5$ or $1/z^6$ from the high energy expansion.
We also demand that the difference in the degree of the polynomial
in the numerator and denominator is less or equal to two.
If one of these conditions is loosened the size of the error
would only slightly increase and the range would
then cover the exact value.
Furthermore one should remember that only information from small and large
external momenta enter into the Pad\'e analysis. The numerical values,
however, are extracted for $q^2=M^2$.
They arise from combinations of different pieces where each one exhibits
singularities at threshold. Thus, in general
one can not expect that an agreement
up to three or more digits with the exact result can be found.

For the construction of the results in Tab.~\ref{tab:12loop}
the same input information was used which will be available at three-loop
order.
At one- and two-loop level, however, more moments can be computed.
In Tab.~\ref{tab:12loop_2} results are shown where terms up to order
$z^8$, respectively, $1/z^8$ are incorporated.
Indeed, the general tendency is a slight reduction in the error
for all colour factors.
{\footnotesize
\begin{table}[t]
\begin{center}
\begin{tabular}{|l|l||r|r|r|r|r|} 
\hline
$n$ & P.A.& $ F $& $ FF $& $ FA $& $ FL $& $ FH $\\
\hline
$7$ & $[6/7]$ & --- &  ${}^\star-0.4954$ &  ${}^\star-3.3596$ &  ${}^\star1.5643$ & --- \\
$7$ & $[6/8]$ &  $-1.0002$ &  ${}^\star-0.4926$ &  ${}^\star-3.4026$ &  $1.5638$ &  ${}^\star-0.1542$ \\
$7$ & $[7/6]$ &  $-1.0003$ &  ${}^\star-0.4953$ &  ${}^\star-3.3603$ &  $1.5641$ &  $-0.1555$ \\
$7$ & $[7/7]$ &  $-1.0002$ &  ${}^\star-0.4925$ &  ${}^\star-3.4088$ &  $1.5638$ &  ${}^\star-0.1541$ \\
$7$ & $[7/8]$ & --- &  $-0.4945$ &  $-3.3722$ & --- &  ${}^\star-0.1517$ \\
$7$ & $[8/6]$ &  $-1.0002$ &  ${}^\star-0.4926$ &  ${}^\star-3.3998$ &  $1.5638$ &  ${}^\star-0.1541$ \\
$7$ & $[8/7]$ &  ${}^\star-1.0003$ & --- & --- &  ${}^\star1.5644$ &  ${}^\star-0.1288$ \\
\hline 
$8$ & $[6/8]$ &  $-1.0002$ &  ${}^\star-0.4930$ &  $-3.3073$ &  $1.5639$ &  $-0.1552$ \\
$8$ & $[7/7]$ &  $-1.0002$ & --- & --- &  $1.5638$ &  $-0.1552$ \\
$8$ & $[7/8]$ & --- &  ${}^\star-0.4945$ & --- & --- &  $-0.1551$ \\
$8$ & $[7/9]$ &  $-1.0001$ &  $-0.4991$ &  ${}^\star-3.3593$ &  $1.5632$ &  $-0.1554$ \\
$8$ & $[8/6]$ &  $-1.0002$ &  ${}^\star-0.4943$ & --- &  $1.5639$ &  $-0.1552$ \\
$8$ & $[8/7]$ &  ${}^\star-1.0003$ &  ${}^\star-0.4936$ & --- &  ${}^\star1.5646$ &  ${}^\star-0.1547$ \\
$8$ & $[8/8]$ &  $-1.0001$ &  $-0.4989$ &  ${}^\star-3.3607$ &  $1.5632$ &  $-0.1554$ \\
$8$ & $[9/7]$ &  $-1.0002$ &  $-0.4993$ &  ${}^\star-3.3594$ &  $1.5633$ &  $-0.1557$ \\
\hline 
\end{tabular}
\caption{\label{tab:12loop_2}
  Results for the one- and two-loop coefficients where terms up to order
  $z^8$, respectively, $1/z^8$ are incorporated.
}
\end{center}
\end{table}
}

As compared to Tab.~\ref{tab:12loop} the situation for FH becomes quite
different.
The inclusion of more low-energy moments leads to significantly
more stable results.
Note, however, that at three-loop order the input data for $n=7$ and $n=8$ are
not available which means that for the diagrams of such type we follow the
strategy outlined above.

At three-loop level the moments are only available up to order $z^6$,
respectively, $1/z^6$. The results obtained with these input data
(cf. Tabs.~\ref{tab:12loop} and~\ref{tab:FHzw})
read~\cite{CheSte99}
\begin{eqnarray}
\begin{array}{lll}
\displaystyle
z_m^{F} \,\,=\,\, -1.0005(8)
\,,
&
\displaystyle
z_m^{FF} \,\,=\,\, -0.49(1)
\,,
&
\displaystyle
z_m^{FA} \,\,=\,\, -3.4(1)
\,,
\\
&
\displaystyle
z_m^{FL} \,\,=\,\, 1.566(5)
\,,
&
z_m^{FH} \,\,=\,\, -0.1553(2)
\,.
\label{eq:zm12num}
\end{array}
\end{eqnarray}
where the error is obtained by doubling the spread of the
different Pad\'e approximants.
The comparison with the exact result results~\cite{GraBroGraSch90}
$\{-1$, $-0.51056$, $-3.33026$, $1.56205$, $-0.15535\}$ shows 
very good agreement --- except for the structure $FF$ where our error estimate
is off by a factor of two.
We will come back to that point below after considering the general-$\xi$
results.

Up to now the discussion was based on Feynman gauge only, which corresponds
to $\xi=0$ in our notation. In the remaining part of this section we
allow for a general gauge parameter and demonstrate how the results of
Eq.~(\ref{eq:zm12num}) can be improved. At one- and two-loop level only
the colour structures $C_F$, $C_F^2$ and $C_FC_A$ develop a $\xi$ dependence.
The proper combination of the self energies
which contributes to the relation between the on-shell and 
$\overline{\rm MS}$ mass evaluated at $z=1$
must be independent of the specific choice of $\xi$.
On the other hand, the expansion terms actually depend on the QCD gauge
parameter, $\xi$.
As we are dealing with a finite number of terms we are left with a
residual $\xi$ dependence in the final result.
It is clear that for extreme values of $\xi$ any predictive power of our
procedure gets lost as the coefficients of $\xi$ become dominant. However,
the final values should be stable against
small variations around $\xi=0$.
Thus it is worth to examine the dependence on $\xi$.

\begin{table}[t]
\begin{center}
\begin{tabular}{cc}
{\tiny
\begin{tabular}{|l|l||r|r|r|} 
\hline
$n$ & P.A.& $ F $& $ FF $& $ FA $\\
\hline
$5$ & $[4/5]$ &  ${}^\star-0.9982$ &  $-0.4644$ &  $-3.4939$ \\
$5$ & $[4/6]$ &  $-0.9989$ &  $-0.4618$ &  $-3.2830$ \\
$5$ & $[5/4]$ &  $-0.9986$ &  $-0.4596$ &  ${}^\star-3.2273$ \\
$5$ & $[5/5]$ &  $-0.9990$ &  $-0.4673$ &  $-3.3145$ \\
$5$ & $[5/6]$ &  $-0.9996$ &  ${}^\star-0.4617$ &  $-3.2547$ \\
$5$ & $[6/4]$ &  $-0.9989$ &  $-0.4629$ &  $-3.2604$ \\
$5$ & $[6/5]$ & --- &  ${}^\star-0.4616$ &  $-3.2570$ \\
\hline 
$6$ & $[4/6]$ &  $-0.9988$ &  ${}^\star-0.4468$ &  $-3.2946$ \\
$6$ & $[5/5]$ &  $-0.9989$ &  $-0.4770$ &  $-3.3044$ \\
$6$ & $[5/6]$ &  $-0.9995$ &  ${}^\star-0.4574$ &  $-3.2613$ \\
$6$ & $[5/7]$ &  $-0.9994$ &  $-0.4721$ &  ${}^\star-3.2672$ \\
$6$ & $[6/4]$ &  $-0.9989$ &  $-0.4670$ &  $-3.2513$ \\
$6$ & $[6/5]$ & --- &  ${}^\star-0.4561$ &  $-3.2675$ \\
$6$ & $[6/6]$ &  $-0.9994$ &  ${}^\star-0.4518$ &  ${}^\star-3.2837$ \\
$6$ & $[7/5]$ &  $-0.9994$ &  $-0.4739$ &  ${}^\star-3.2628$ \\
\hline 
\end{tabular}
}
&
{\tiny
\begin{tabular}{|l|l||r|r|r|} 
\hline
$n$ & P.A.& $ F $& $ FF $& $ FA $\\
\hline
$5$ & $[4/5]$ & --- &  $-0.5105$ & --- \\
$5$ & $[4/6]$ &  $-1.0000$ &  $-0.5040$ &  $-3.3521$ \\
$5$ & $[5/4]$ &  $-1.0000$ &  $-0.5105$ &  $-3.3531$ \\
$5$ & $[5/5]$ &  $-1.0000$ &  ${}^\star-0.5209$ &  $-3.3499$ \\
$5$ & $[5/6]$ &  $-1.0000$ &  ${}^\star-0.5100$ &  ${}^\star-3.3521$ \\
$5$ & $[6/4]$ &  $-1.0000$ &  $-0.5055$ &  $-3.3517$ \\
$5$ & $[6/5]$ &  ${}^\star-1.0000$ &  ${}^\star-0.5099$ &  ${}^\star-3.3521$ \\
\hline 
$6$ & $[4/6]$ &  $-1.0000$ &  $-0.5070$ &  ${}^\star-3.3576$ \\
$6$ & $[5/5]$ &  $-1.0000$ &  ${}^\star-0.5119$ & --- \\
$6$ & $[5/6]$ &  $-1.0000$ &  ${}^\star-0.5093$ &  ${}^\star-3.3540$ \\
$6$ & $[5/7]$ &  $-1.0000$ &  ${}^\star-0.5114$ &  $-3.3472$ \\
$6$ & $[6/4]$ &  $-1.0000$ &  $-0.5083$ &  $-3.3497$ \\
$6$ & $[6/5]$ &  ${}^\star-1.0000$ &  ${}^\star-0.5092$ &  ${}^\star-3.3545$ \\
$6$ & $[6/6]$ &  $-1.0000$ &  ${}^\star-0.5112$ &  ${}^\star-3.3557$ \\
$6$ & $[7/5]$ &  $-1.0000$ &  ${}^\star-0.5115$ &  $-3.3464$ \\
\hline 
\end{tabular}
}
\end{tabular}
\caption{\label{tab:xi_1}
  Results for the structures $F$, $FF$ and $FA$ where for the gauge parameter
  the values $\xi=-5$ (left table) and $\xi=-2$ (right table) have been
  adopted. 
}
\end{center}
\end{table}
\begin{table}[t]
\begin{center}
\begin{tabular}{cc}
{\tiny
\begin{tabular}{|l|l||r|r|r|} 
\hline
$n$ & P.A.& $ F $& $ FF $& $ FA $\\
\hline
$5$ & $[4/5]$ &  ${}^\star-1.0021$ &  $-0.4344$ &  $-3.3930$ \\
$5$ & $[4/6]$ &  $-1.0013$ &  ${}^\star-0.4176$ &  ${}^\star-3.4638$ \\
$5$ & $[5/4]$ &  $-1.0017$ &  $-0.4209$ &  $-3.4105$ \\
$5$ & $[5/5]$ &  $-1.0013$ &  $-0.4473$ &  $-3.3778$ \\
$5$ & $[5/6]$ &  $-1.0005$ &  ${}^\star-0.4212$ &  ${}^\star-3.4175$ \\
$5$ & $[6/4]$ &  $-1.0013$ &  $-0.4286$ &  $-3.4050$ \\
$5$ & $[6/5]$ & --- &  ${}^\star-0.4206$ &  ${}^\star-3.4195$ \\
\hline 
$6$ & $[4/6]$ &  $-1.0014$ & --- &  $-3.3301$ \\
$6$ & $[5/5]$ &  $-1.0013$ &  $-0.4597$ & --- \\
$6$ & $[5/6]$ &  $-1.0006$ &  ${}^\star-0.4099$ &  ${}^\star-3.4389$ \\
$6$ & $[5/7]$ &  $-1.0007$ &  $-0.4397$ &  ${}^\star-3.3990$ \\
$6$ & $[6/4]$ &  $-1.0014$ &  $-0.4353$ &  ${}^\star-3.3972$ \\
$6$ & $[6/5]$ & --- &  ${}^\star-0.4024$ &  ${}^\star-3.4692$ \\
$6$ & $[6/6]$ &  $-1.0007$ &  ${}^\star-0.2736$ &  ${}^\star-3.2798$ \\
$6$ & $[7/5]$ &  $-1.0008$ &  $-0.4444$ &  $-3.3914$ \\
\hline 
\end{tabular}
}
&
{\tiny
\begin{tabular}{|l|l||r|r|r|} 
\hline
$n$ & P.A.& $ F $& $ FF $& $ FA $\\
\hline
$5$ & $[4/5]$ &  ${}^\star-1.0039$ &  $-0.3230$ &  $-3.3834$ \\
$5$ & $[4/6]$ &  $-1.0024$ & --- &  ${}^\star-3.4196$ \\
$5$ & $[5/4]$ &  $-1.0030$ &  $-0.2620$ &  $-3.3970$ \\
$5$ & $[5/5]$ &  $-1.0022$ &  $-0.3700$ &  $-3.3713$ \\
$5$ & $[5/6]$ &  $-1.0009$ &  ${}^\star-0.2110$ &  ${}^\star-3.4006$ \\
$5$ & $[6/4]$ &  $-1.0023$ & --- &  $-3.3917$ \\
$5$ & $[6/5]$ & --- &  ${}^\star-0.1970$ &  ${}^\star-3.4017$ \\
\hline 
$6$ & $[4/6]$ &  $-1.0025$ &  $-0.4670$ &  $-3.3204$ \\
$6$ & $[5/5]$ &  $-1.0024$ &  $-0.3853$ &  $-3.3643$ \\
$6$ & $[5/6]$ &  $-1.0010$ &  ${}^\star-0.0740$ &  ${}^\star-3.4146$ \\
$6$ & $[5/7]$ &  $-1.0013$ &  ${}^\star-0.2749$ &  ${}^\star-3.3856$ \\
$6$ & $[6/4]$ &  $-1.0025$ &  ${}^\star-0.3001$ &  ${}^\star-3.3857$ \\
$6$ & $[6/5]$ & --- & --- &  ${}^\star-3.4305$ \\
$6$ & $[6/6]$ &  $-1.0012$ &  ${}^\star-0.5118$ & --- \\
$6$ & $[7/5]$ &  $-1.0014$ &  $-0.3122$ &  $-3.3803$ \\
\hline 
\end{tabular}
}
\end{tabular}
\caption{\label{tab:xi_2}
  Results for the structures $F$, $FF$ and $FA$ where for the gauge parameter
  the values $\xi=+2$ (left table) and $\xi=+5$ (right table) have been
  adopted. 
}
\end{center}
\end{table}

In Tabs.~\ref{tab:xi_1} and~\ref{tab:xi_2} the results are shown for the
$F$, $FF$ and $FA$ structures where the values $\xi=\pm5$
and $\xi=\pm2$ have been adopted\footnote{Despite the fact
  that for $\xi>1$ the generating functional is in principle not defined we
  decided to choose this range for the gauge parameter.}.
The one-loop result is quite stable against the variation of $\xi$
and even for $\xi=\pm5$ the deviations to the exact result is small.
Also the results for $C_AC_F$ exhibit a quite small variation
for the different values of $\xi$.
Only the structure $FF$ shows larger deviations from the exact value if $\xi$
is varied. Especially the average value for $\xi=+5$ is off by roughly 40\%.
However, one also has to take into account that in this case the procedure
seems to be less stable and the errors are larger.
A closer look into the results for the expansion terms shows that for this
choice the coefficients of the $z^n$ terms are already dominated by the
$\xi$-dependent terms whereas, for instance, for $\xi=-5$ this is not the
case. 

For all colour structures the best results are obtained for the choice
$\xi=-2$. This can be understood with the help of the following
considerations. 
As is well known~\cite{Landau54} the QED fermion propagator get infra-red
finite (in the leading log approximation) in a particular gauge,
namely for $\xi=-2$ in our notation (cf. Eq.~(\ref{eq:gluprop})).
Thus, it is not surprising that for this particular choice our approximate
results for $z_m$ demonstrate better convergence to the known ones.
It also suggests that the use of the same gauge condition ($\xi=-2$) should
lead to better results at least for the QED parts of $z_m$ at three-loop
order. This will be discussed in detail in Section~\ref{sec:3loop}.

\begin{figure}[t]
  \begin{center}
    \begin{tabular}{c}
      \leavevmode
      \epsfxsize=14.cm
      \epsffile[90 400 510 720]{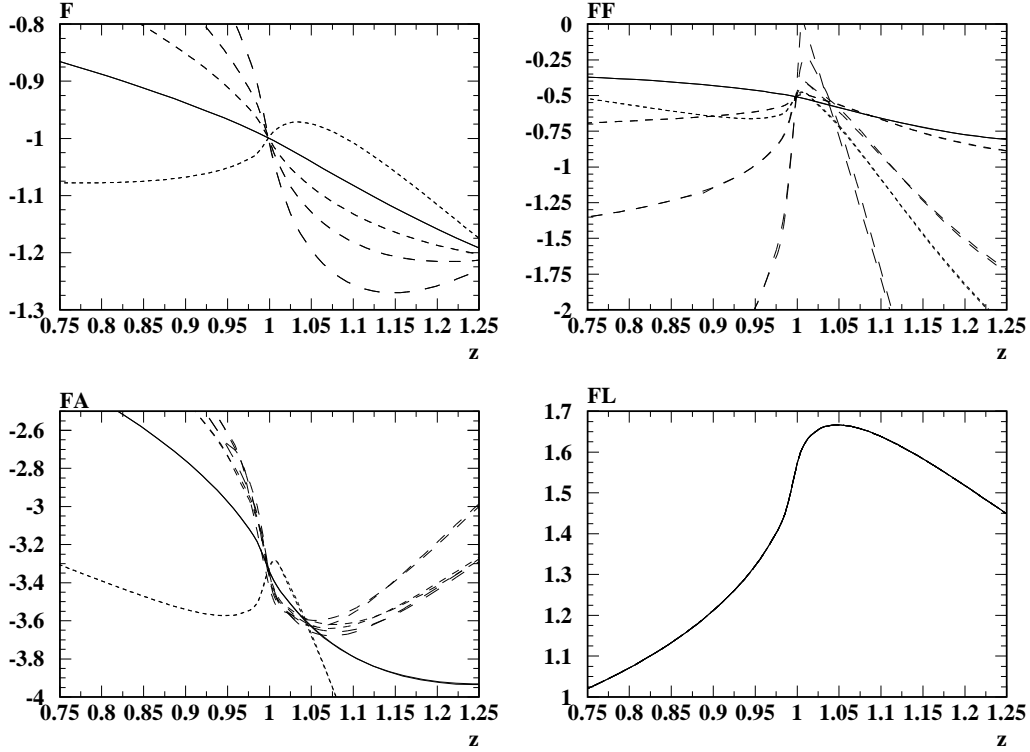}
    \end{tabular}
  \end{center}
  \caption{\label{fig:xidep}
    $z$ dependence of $-g(z)$ for the structures $F$, $FF$ and $FA$ for
    different values of $\xi$. The solid curves belong to $\xi=-2$.
    The dashed curves correspond to $\xi=-5,0,+2,+5$ (from short to long
    dashes).
    }
\end{figure}
In Fig.~\ref{fig:xidep}
the structures $F$, $FF$ and $FA$ of the
function $-g(z)$ are plotted for five different values of $\xi$. 
In each figure ten different curves are plotted --- two different Pad\'e
results for each value of $\xi$. 
Whereas for $z\not=1$ the single curves show a quite different behaviour
there is the tendency to converge to the same value for $z=1$
which corresponds to the respective coefficient of $z_m$.
For comparison also the result for $FL$ is shown in Fig.~\ref{fig:xidep}
which has no $\xi$ dependence at all and which agrees with the exact result
within 0.3\%. 
In the case of $F$ all curves are smooth at the point $z=1$
which is reflected in the impressive stability shown in
Tabs.~\ref{tab:12loop},~\ref{tab:xi_1} and~\ref{tab:xi_2}.
On the contrary, for most values of $\xi$ the two-loop structures
$FF$ and $FA$ exhibit a kink for $z=1$.
In the case of $FF$ even the sign of the derivative is different for $z<1$ and
$z>1$ and a relatively strong variation around $z=1$ is observed.
A closer look shows that the value $\xi=-2$ obviously represents an
exception. For this choice the curves exhibit the smoothest
behaviour around $z=1$. This is also reflected in the numbers presented in
Tab.~\ref{tab:xi_1} which are most stable for $\xi=-2$.
Actually the structure $FF$ has stronger singularities at threshold for
$\xi=0$ than the other coefficients which explains the deviation
from the exact result in Eq.~(\ref{eq:zm12num}).
Thus it is suggestive to choose $\xi=-2$ in order to extract the
results:
\begin{eqnarray}
\begin{array}{lll}
\displaystyle
z_m^{F} \,\,=\,\, -1.0000(0)
\,,
&
\displaystyle
z_m^{FF} \,\,=\,\, -0.51(2)
\,,
&
\displaystyle
z_m^{FA} \,\,=\,\, -3.35(1)
\,,
\label{eq:zm12num2}
\end{array}
\end{eqnarray}
where the errors were obtained in the same way as before.
We would like to stress that there is an impressive agreement with the exact
results (cf. Eq.~(\ref{eq:zm12loop})).

Despite the fact that the uncertainties in Eqs.~(\ref{eq:zm12num2}) are quite
small the total error would be significantly overestimated if one would naively
add the contributions from the individual colour factors. Instead it is more
promising to add the moments in a first step (of course, taking into account
the correct colour factors) and performing the conformal mapping and Pad\'e
approximation afterwards. The results for the quantity
\begin{eqnarray}
z_m^{(2),light}(n_l) &=& C_F^2 z_m^{FF} + C_FC_A z_m^{FA} + C_FTn_l z_m^{FL}
\,
\end{eqnarray}
for different choices of $n_l$ can be found in
Tabs.~\ref{tab:2lfixnl1} and~\ref{tab:2lfixnl2} where
the gauge parameter has been fixed to $\xi=0$ and $\xi=-2$, respectively.
In both cases the spread among the different results is very
small.
Using again twice the spread of the different Pad\'e approximants as an
estimate for the error we get for $\xi=0$
\begin{eqnarray}
  z_m^{(2),light}(0)=-14.4(6)
  \,,
  &
  z_m^{(2),light}(1)=-13.3(6)
  \,,
  &
  z_m^{(2),light}(2)=-12.3(4)
  \,,
  \nonumber\\
  z_m^{(2),light}(3)=-11.2(5)
  \,,
  &
  z_m^{(2),light}(4)=-10.2(3)
  \,,
  &
  z_m^{(2),light}(5)=-9.1(3)
  \,.
\end{eqnarray}
Good agreement with the exact results, which are listed in the last column of
the tables, is obtained.
For $\xi=-2$ the following results can be extracted
\begin{eqnarray}
  z_m^{(2),light}(0)=-14.3(2)
  \,,
  &
  z_m^{(2),light}(1)=-13.2(2)
  \,,
  &
  z_m^{(2),light}(2)=-12.2(2)
  \,,
  \nonumber\\
  z_m^{(2),light}(3)=-11.1(1)
  \,,
  &
  z_m^{(2),light}(4)=-10.07(3)
  \,,
  &
  z_m^{(2),light}(5)=-9.02(1)
  \,.
\end{eqnarray}
Here the error is slightly reduced and is well below 2\%.

{\footnotesize
\begin{table}[ht]
\begin{center}
\begin{tabular}{|l|l||r|r|r|r|r|r|} 
\hline
$n$ & P.A.& $ n_l=0 $& $ n_l=1 $& $ n_l=2 $& $ n_l=3 $& $ n_l=4 $& $ n_l=5 $\\
\hline
$5$ & $[4/5]$ &  $-14.3431$ &  $-13.2932$ &  $-12.2433$ &  $-11.1933$ &  $-10.1432$ &  $-9.0931$ \\
$5$ & $[4/6]$ &  $-14.2673$ &  $-13.2188$ &  $-12.1700$ &  $-11.1206$ &  $-10.0703$ &  $-9.0183$ \\
$5$ & $[5/4]$ &  $-14.3816$ &  $-13.3305$ &  $-12.2791$ &  $-11.2273$ &  $-10.1750$ &  $-9.1221$ \\
$5$ & $[5/5]$ &  $-14.3188$ &  $-13.2702$ &  $-12.2216$ &  $-11.1730$ &  $-10.1244$ &  $-9.0758$ \\
$5$ & $[5/6]$ &  ${}^\star-14.7124$ &  ${}^\star-13.5777$ &  ${}^\star-12.4638$ &  ${}^\star-11.3641$ &  ${}^\star-10.2745$ &  ${}^\star-9.1925$ \\
$5$ & $[6/4]$ &  ${}^\star-14.4299$ &  ${}^\star-13.3703$ &  ${}^\star-12.3109$ &  ${}^\star-11.2517$ &  ${}^\star-10.1927$ &  ${}^\star-9.1340$ \\
$5$ & $[6/5]$ & --- & --- & --- &  ${}^\star-11.4712$ &  ${}^\star-10.3346$ &  ${}^\star-9.2264$ \\
\hline 
$6$ & $[4/6]$ &  $-14.2802$ &  $-13.2344$ &  $-12.1885$ &  $-11.1425$ &  $-10.0963$ &  $-9.0499$ \\
$6$ & $[5/5]$ &  $-14.3149$ &  $-13.2662$ &  $-12.2176$ &  $-11.1689$ &  $-10.1203$ &  $-9.0717$ \\
$6$ & $[5/7]$ &  ${}^\star-14.5630$ &  ${}^\star-13.4629$ &  ${}^\star-12.3745$ &  ${}^\star-11.2942$ &  ${}^\star-10.2200$ &  ${}^\star-9.1502$ \\
$6$ & $[6/4]$ &  ${}^\star-14.3676$ &  ${}^\star-13.3163$ &  ${}^\star-12.2647$ &  ${}^\star-11.2130$ &  ${}^\star-10.1610$ &  ${}^\star-9.1088$ \\
$6$ & $[6/5]$ &  $-14.2190$ &  $-13.1684$ &  $-12.1164$ &  $-11.0623$ &  $-10.0043$ &  $-8.9382$ \\
$6$ & $[6/6]$ &  ${}^\star-14.2867$ &  ${}^\star-13.2389$ &  ${}^\star-12.1910$ &  ${}^\star-11.1431$ &  ${}^\star-10.0951$ &  ${}^\star-9.0469$ \\
$6$ & $[7/5]$ &  $-14.3713$ &  $-13.3191$ &  $-12.2668$ &  $-11.2142$ &  $-10.1613$ &  $-9.1082$ \\
\hline 
&exact& $-14.2287$&$-13.1874$&$-12.146$&$-11.1046$&$-10.0633$&$-9.02188$\\
\hline
\end{tabular}
\caption{\label{tab:2lfixnl1}Results for $z_m^{(2),light}(n_l)$ for different
  values of $n_l$. The gauge parameter has been fixed to $\xi=0$.
}
\end{center}
\end{table}
}

{\footnotesize
\begin{table}[ht]
\begin{center}
\begin{tabular}{|l|l||r|r|r|r|r|r|} 
\hline
$n$ & P.A.& $ n_l=0 $& $ n_l=1 $& $ n_l=2 $& $ n_l=3 $& $ n_l=4 $& $ n_l=5 $\\
\hline
$5$ & $[4/5]$ &  $-14.2809$ &  $-13.2304$ &  $-12.1797$ &  $-11.1284$ &  $-10.0754$ &  ${}^\star-9.0136$ \\
$5$ & $[4/6]$ &  $-14.1864$ & --- & --- &  ${}^\star-11.1492$ &  ${}^\star-10.0769$ &  $-9.0194$ \\
$5$ & $[5/4]$ &  $-14.3044$ &  $-13.2494$ &  $-12.1933$ &  $-11.1360$ &  $-10.0771$ &  $-9.0164$ \\
$5$ & $[5/5]$ &  $-14.2670$ &  $-13.2184$ &  $-12.1698$ &  $-11.1212$ &  $-10.0726$ &  $-9.0245$ \\
$5$ & $[5/6]$ &  ${}^\star-14.3356$ &  ${}^\star-13.2670$ &  ${}^\star-12.2014$ &  ${}^\star-11.1382$ &  ${}^\star-10.0769$ &  ${}^\star-9.0173$ \\
$5$ & $[6/4]$ &  ${}^\star-14.3073$ &  ${}^\star-13.2492$ &  $-12.1913$ &  $-11.1336$ &  $-10.0760$ &  $-9.0186$ \\
$5$ & $[6/5]$ &  ${}^\star-14.3458$ &  ${}^\star-13.2720$ &  ${}^\star-12.2035$ &  ${}^\star-11.1387$ &  ${}^\star-10.0769$ &  ${}^\star-9.0173$ \\
\hline 
$6$ & $[4/6]$ &  $-14.2460$ &  $-13.1986$ &  $-12.1504$ &  $-11.0983$ &  ${}^\star-10.0940$ &  $-9.0191$ \\
$6$ & $[5/5]$ &  $-14.2625$ &  $-13.2139$ &  $-12.1653$ &  $-11.1167$ &  $-10.0677$ &  $-9.0202$ \\
$6$ & $[5/6]$ & --- &  ${}^\star-13.3253$ &  ${}^\star-12.2248$ &  ${}^\star-11.1461$ &  ${}^\star-10.0785$ &  $-9.0172$ \\
$6$ & $[5/7]$ &  ${}^\star-14.3106$ &  ${}^\star-13.2490$ &  ${}^\star-12.1892$ &  ${}^\star-11.1309$ &  $-10.0737$ &  ${}^\star-9.0177$ \\
$6$ & $[6/4]$ &  ${}^\star-14.2928$ &  ${}^\star-13.2394$ &  ${}^\star-12.1853$ &  ${}^\star-11.1306$ &  $-10.0748$ &  $-9.0179$ \\
$6$ & $[6/5]$ & --- & --- &  ${}^\star-12.2729$ &  ${}^\star-11.1535$ &  ${}^\star-10.0790$ &  $-9.0175$ \\
$6$ & $[6/6]$ &  ${}^\star-14.2375$ &  ${}^\star-13.1871$ &  ${}^\star-12.1322$ & --- &  ${}^\star-10.0819$ &  ${}^\star-9.0177$ \\
$6$ & $[7/5]$ &  $-14.2909$ &  $-13.2370$ &  $-12.1827$ &  $-11.1280$ &  $-10.0731$ &  ${}^\star-9.0177$ \\
\hline 
&exact& $-14.2287$&$-13.1874$&$-12.146$&$-11.1046$&$-10.0633$&$-9.02188$\\
\hline
\end{tabular}
\caption{\label{tab:2lfixnl2}Results for $z_m^{(2),light}(n_l)$ for different 
  values of $n_l$. The gauge parameter has been fixed to $\xi=-2$.
}
\end{center}
\end{table}
}

In summary, very good agreement with the exact results for the different
coefficients of $z_m$ are observed at one- and two-loop order.
The results obtained from the fifth and sixth low-energy moment and a
high-energy expansion up to ${\cal O}(1/z^6)$ suggests that also at order
$\alpha_s^3$ it is possible to get reliable results.
The analysis for different values of $\xi$ showed that the choice $\xi=-2$
would be the preferable one. However, also $\xi=0$ leads to very good
results. At three-loop order the calculation for general $\xi$
is only possible for some of the diagrams, namely the ones which are already
present in QED. For those diagrams, which have stronger singularities for
$z=1$, $\xi=-2$ will be adopted.
In the remaining cases the results are obtained for $\xi=0$.
In the approach where the moments are added in a first step our method
provides for $\xi=0$ results which are in very good agreement with the exact
ones. 


\section{\label{sec:master}Master integrals}

Before we go on to the discussion of the three-loop results the technology
behind the practical calculation shall be discussed in more detail.
The high-energy expansion can be reduced to the evaluation of massless
propagator-type diagrams up to three loops and one- and two-loop vacuum
graphs. The relevant tools are available since quite some time (see,
e.g.,~\cite{HarSte98}).
The technology needed in the limit $z\to0$, however, has never
been applied to real processes. Furthermore the discussion in the literature
is not complete. In this Section we review the knowledge about vacuum graphs
and close the gaps.

\begin{figure}[t]
  \begin{center}
    \begin{tabular}{c}
      \leavevmode
      \epsfxsize=3.cm
      \epsffile[170 290 415 530]{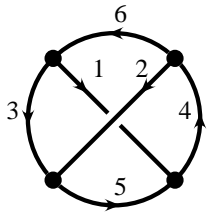}
    \end{tabular}
  \end{center}
  \caption{\label{fig:master}
    Generic three-loop bubble diagram. In general each line may be massless
    or carry mass $M$.}
\end{figure}
The computation of the three-loop diagrams in the limit of vanishing external
momentum reduces to the evaluation of vacuum graphs
as shown in Fig.~\ref{fig:master} 
with one dimensionful scale given by the mass, $M$.
In principle each line may be massless or carry mass $M$.
The general strategy to compute such kind of diagrams is based on the
integration-by-parts method~\cite{CheTka81}. It provides recurrence relations
which are used in order to express complicated integrals in terms of simple
ones and a small set of a few so-called master integrals for which a real
integration is necessary. Those master integrals, however, only have to be
computed once and forever.

\begin{figure}[t]
  \begin{center}
    \begin{tabular}{c}
      \leavevmode
      \epsfxsize=10.cm
      \epsffile[120 500 460 720]{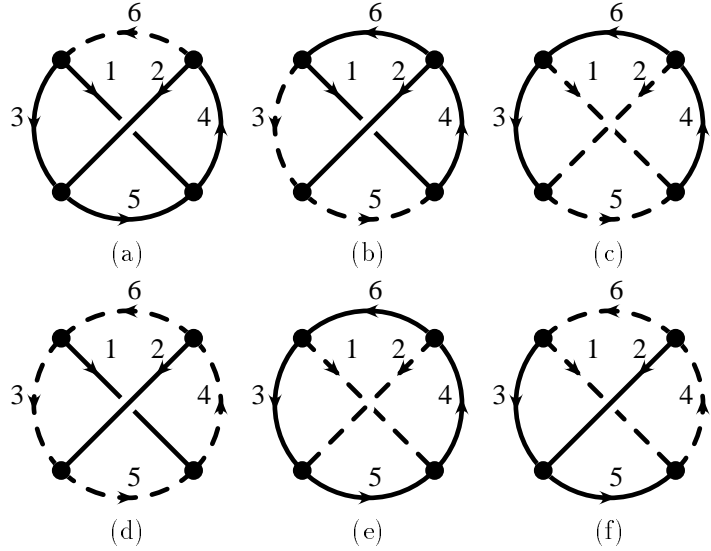}
    \end{tabular}
  \end{center}
  \caption{\label{fig:master1}
    Master diagrams corresponding to the expression in
    Eq.~(\ref{eq:master1}). The full lines carry mass $M$ whereas the dashed
    ones are massless.}
\end{figure}

\begin{figure}[t]
  \begin{center}
    \begin{tabular}{ccc}
      \leavevmode
      \epsfxsize=3.cm
      \epsffile[170 290 415 530]{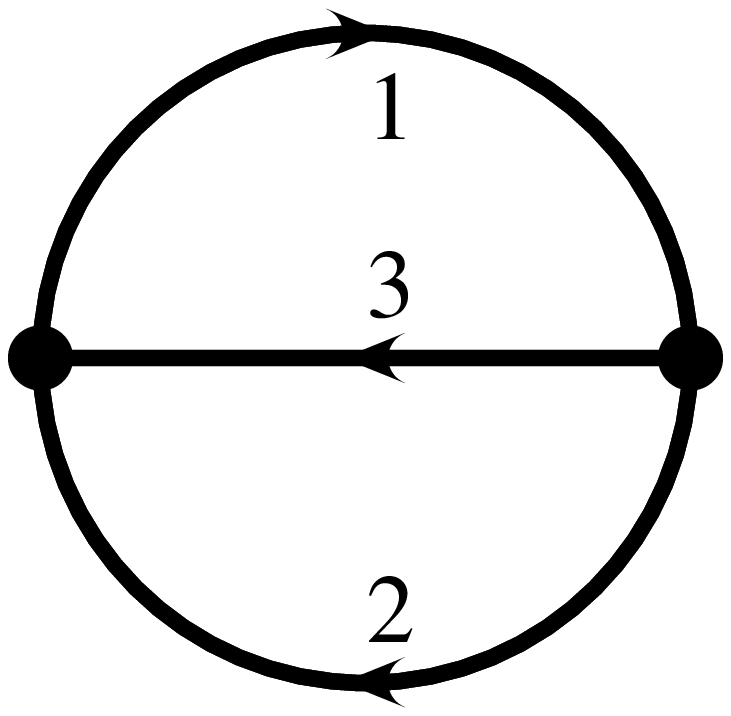}
      &
      \leavevmode
      \epsfxsize=3.cm
      \epsffile[170 290 415 530]{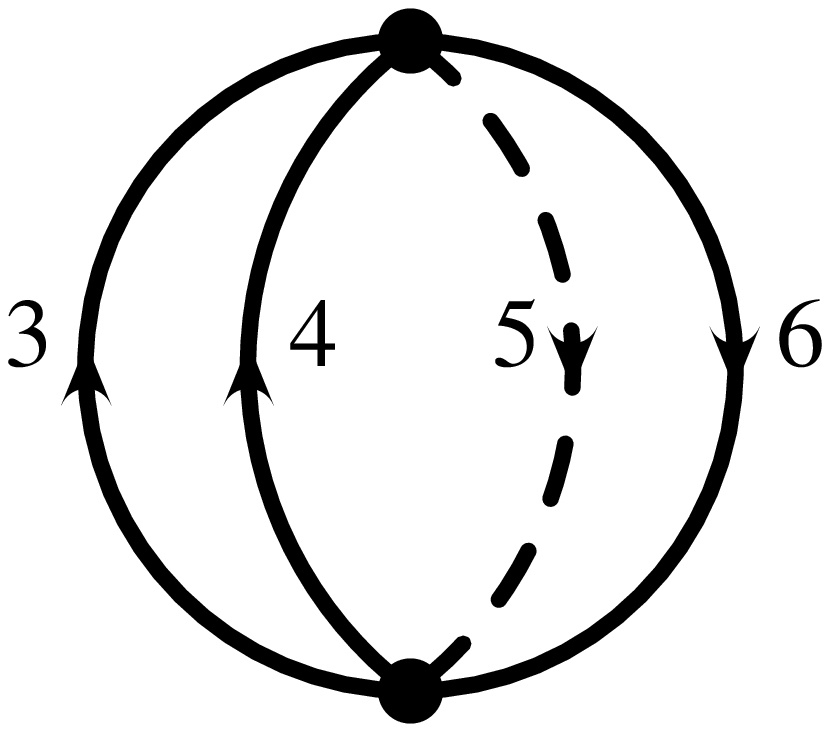}
      &
      \leavevmode
      \epsfxsize=3.cm
      \epsffile[170 290 415 530]{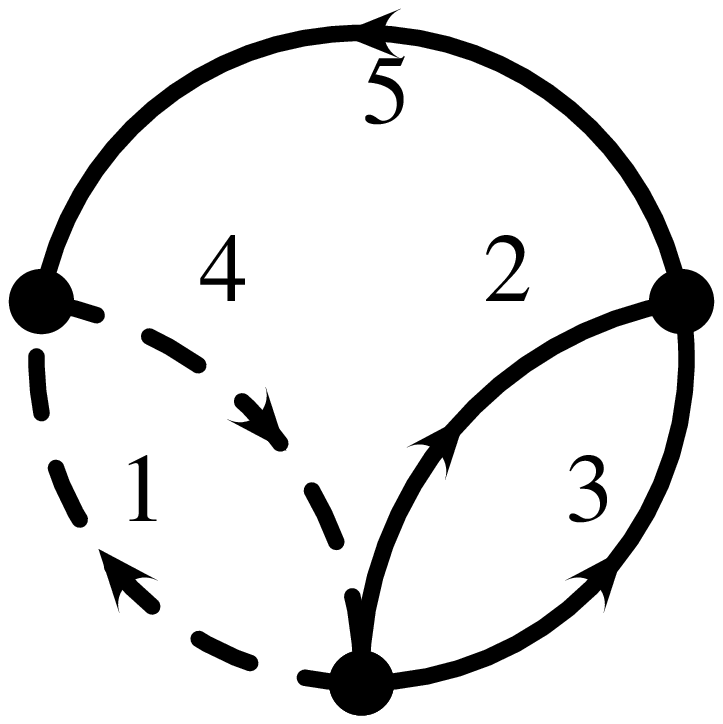}
      \\
      (a) & (b) & (c)
    \end{tabular}
  \end{center}
  \caption{\label{fig:master2}
    Two- and three-loop master diagrams corresponding to the expressions in
    Eqs.~(\ref{eq:master2}) and~(\ref{eq:master2l}).
    The full lines carry mass $M$ whereas the dashed
    ones are massless.}
\end{figure}

The recurrence relations and master integrals which are necessary for the
computation of ${\cal O}(\alpha_s^2)$ corrections to the $\gamma$ and $Z$
boson propagator have been considered in~\cite{Bro92}.
In this case only one difficult\footnote{Here we mean those master integrals
  which cannot be solved by successive application of one- and two-loop
  formulae.}
integral had to be solved. It is shown in Fig.~\ref{fig:master1}(e).
In~\cite{Avdrho} the technique has been extended to the case of the $W$
boson. Altogether only three difficult master integrals are necessary for the
evaluation of the gauge boson self energies. In addition to the one in
Fig.~\ref{fig:master1}(e) also the diagrams in Fig.~\ref{fig:master1}(c)
and Fig.~\ref{fig:master2}(b) have to be taken into account.

For the fermion propagator the variety of integrals to be computed at
three-loop level is larger as there are more choices whether a certain line of
the three-loop tadpole diagram shown in Fig.~\ref{fig:master}
may be massive or massless.
Let us collect all master integrals which are needed for the computation of
the moments for the fermion propagator. It is convenient to introduce the
following function
\begin{eqnarray}
  \lefteqn{M_{n_1,n_2,n_6}(M_1,\ldots,M_6) = }
  \\&&\mbox{}
  \int
  \frac{{\rm d}^Dp\,{\rm d}^Dk\,{\rm d}^Dl}{(M^2\pi)^{3D/2}}
  \frac{M^{6+2n_1+2n_2+2n_6}}{
    (p_1^2+M_1^2)^{n_1}
    (p_2^2+M_2^2)^{n_2}
    (p_3^2+M_3^2)
    (p_4^2+M_4^2)
    (p_5^2+M_5^2)
    (p_6^2+M_6^2)^{n_6}
    }
  \nonumber
  \,,
  \label{eq:master_def}
\end{eqnarray}
where the masses $M_i$ are either equal to $M$ or zero.
$D=4-2\varepsilon$ is the space-time dimension.
The factors of $M^2$ have been introduced in order to make the function
$M(\ldots)$ dimensionless.
The momenta $p_i$ are linear combinations of the integration momenta
$p$, $k$ and $l$.
In~\cite{Avd97} the recurrence relations for those mass decompositions were
derived which were not discussed in~\cite{Bro92}.
They can be used in order to arrive at the master integrals
pictured in Fig.~\ref{fig:master1}.
The results can be found in~\cite{Bro92,Bro98}\footnote{Following standard
  $\overline{\rm MS}$ practice $\gamma_E$ and $\ln4\pi$ is discarded in the
  in the explicit expressions for the integrals. In practice this means that
  we multiply 
  with a factor $e^{\varepsilon\ln(4\pi)+\varepsilon\gamma_E}$ for each loop.}
\begin{eqnarray}
  M_{111}(M,M,M,M,M,0) &=&
  \frac{2\zeta_3}{\varepsilon}
  + 6\zeta_3-\frac{469}{27}\zeta_4
  +\frac{8}{3}\left[{\rm Cl}_2\left(\frac{\pi}{3}\right)\right]^2
  \nonumber\\&&\mbox{}
  -16\sum_{m>n>0} \frac{(-1)^m \cos(2\pi n/3)}{m^3n}
  \nonumber\\&\approx&
  \frac{2\zeta_3}{\varepsilon}
  -8.2168598175087380629133983386010858249695
  \,,
  \nonumber\\
  M_{111}(M,M,0,M,0,M) &=&
  \frac{2\zeta_3}{\varepsilon} 
  + 6\zeta_3 - \frac{77}{12}\zeta_4 - 
  6\left[{\rm Cl}_2\left(\frac{\pi}{3}\right)\right]^2
  \,,
  \nonumber\\
  M_{111}(0,0,M,M,0,M) &=&
  \frac{2\zeta_3}{\varepsilon}
  + 6\zeta_3 - \frac{15}{4}\zeta_4 - 
  6\left[{\rm Cl}_2\left(\frac{\pi}{3}\right)\right]^2
  \,,
  \nonumber\\
  M_{111}(M,M,0,0,0,0) &=&
  \frac{2\zeta_3}{\varepsilon}
  + 6\zeta_3 - \frac{21}{2}\zeta_4 - 4 \zeta_2\ln^2 2
  + \frac{2}{3}\ln^4 2 
  + 16 {\rm Li}_4\left(\frac{1}{2}\right)
  \,,
  \nonumber\\
  M_{111}(0,0,M,M,M,M) &=&
  \frac{2\zeta_3}{\varepsilon}
  + 6\zeta_3 - 22\zeta_4 - 8 \zeta_2\ln^2 2
  + \frac{4}{3}\ln^4 2 
  + 32 {\rm Li}_4\left(\frac{1}{2}\right)
  \,,
  \nonumber\\
  M_{111}(0,M,M,0,M,0) &=&
  \frac{2\zeta_3}{\varepsilon}
  + 6\zeta_3 - \frac{11}{2}\zeta_4  - 
  4\left[{\rm Cl}_2\left(\frac{\pi}{3}\right)\right]^2
  \,,
  \label{eq:master1}
\end{eqnarray}
where $\zeta_n$ is Rieman's zeta function with the values
$\zeta_2=\pi^2/6$, $\zeta_3\approx 1.20205690$ and $\zeta_4=\pi^4/90$.
${\rm Cl}_2(x)$ is the Clausen function and ${\rm Li}_4(x)$ represents the
quadrilogarithm.
The order of the expressions listed in Eq.~(\ref{eq:master1})
agrees with the one of Fig.~\ref{fig:master1}(a)--(f).
Note that the case where all six lines are massive does not occur for the
three-loop QCD corrections to the quark propagator.

It is interesting to note that the results corresponding to
Figs.~\ref{fig:master1}(a) and~(e) can be obtained from the
${\cal O}(\varepsilon)$ part of the diagrams where the index of
one of the massless lines is reduced to zero using the
integration-by-parts technique.
The resulting topology leads to much simpler expressions as
it can essentially be written as a product of two one-loop integrals with an
additional integration over the corresponding external momentum.
For the other four cases the topologies presented in
Fig.~\ref{fig:master1} have to be considered in order to reproduce the
results of Eq.~(\ref{eq:master1}).
The reason for this is that the
absence of any of the lines immediately leads to (in general simpler)
integrals which at the end result in different master integrals.

In the practical application of the recurrence relations it turns out that
except the master integrals of Eq.~(\ref{eq:master1}) also the one
of Fig.~\ref{fig:master2}(c) is needed. Furthermore
the ${\cal O}(D-4)$ parts of the diagram pictured in Fig.~\ref{fig:master2}(b)
is required\footnote{For some diagrams it is necessary to
  know the ${\cal O}(D-4)$ 
  part as the recurrence relations may contain artificial $1/(D-4)$ poles.}.
These pieces will be provided in the following.

For their practical computation it is convenient to consider
integrals which are less divergent.
This is achieved by doubling some of the propagators.
In the case of Fig.~\ref{fig:master2}(b) we double each of the massive
propagators and for Fig.~\ref{fig:master2}(c) the propagators
corresponding to line 1, 2 and 4 are raised to power two.
This leads us to ultra-violet convergent integrals.
The infra-red divergence introduced in the latter case does not spoil
the evaluation. It manifests itself as as single $1/\varepsilon$ pole in the
final result.
In a next step we introduce Feynman parameters
and thus combine the different denominators. Then the momentum
integrations can be carried out and one ends up with two-
(Fig.~\ref{fig:master2}(b)), respectively,
three-dimensional (Fig.~\ref{fig:master2}(c))
integral representations.
In each case the leading terms of order $1/\varepsilon$, respectively,
$\varepsilon^0$ can be used in
order to check the constants which are already known.
The expansion
up to next-to-leading order in $\varepsilon$ provides the structures
we are interested in. 
In both cases one ends up with a one-dimensional integration
where the most difficult integrals look like:
\begin{eqnarray}
\int_0^1 {\rm d}x \frac{x^n\ln(x)\ln(1-x+x^2)}{(1-x+x^2)^2}
\,,\qquad n=0,1,\ldots
\,.
\end{eqnarray}
The integration can be performed analytically and finally leads to the
following results for the master integrals:
\begin{eqnarray}
  M_{001}(0,0,M,M,0,M) &=&
  \frac{1}{\varepsilon^3}
  +\frac{15}{4\varepsilon^2}
  +\frac{1}{\varepsilon}\left(\frac{65}{8}+\frac{3}{2}\zeta_2\right)
  +\frac{135}{16} + \frac{45}{8}\zeta_2 
  - \zeta_3
  \nonumber\\&&\mbox{}
  +3\sqrt{3}{\rm Cl}_2\left(\frac{\pi}{3}\right)
  +\varepsilon\left(
    - \frac{763}{32}
    - \frac{9\sqrt{3}\pi}{16}\ln^2 3
    - \frac{35\sqrt{3}\pi^3}{48}
  \right.\nonumber\\&&\left.\mbox{}
    + \frac{195}{16}\zeta_2
    - \frac{15}{4}\zeta_3
    + \frac{57}{16}\zeta_4
    + \frac{45\sqrt{3}}{2} {\rm Cl}_2\left(\frac{\pi}{3}\right)
  \right.\nonumber\\&&\left.\mbox{}
    - 27\sqrt{3} {\rm Im}\left[ {\rm
        Li}_3\left(\frac{e^{-i\pi/6}}{\sqrt{3}}\right)\right] 
  \right)
\,,
\nonumber\\
M_{110}(0,M,M,0,M,0) &=&
  -\frac{2}{3\varepsilon^3}
  -\frac{11}{3\varepsilon^2}
  +\frac{1}{\varepsilon}\left(
    -14
    -2\zeta_2
    +2\sqrt{3}{\rm Cl}_2\left(\frac{\pi}{3}\right)
  \right)
  \nonumber\\&&\mbox{}
  - \frac{139}{3} 
  - \frac{\sqrt{3}\pi}{8} \ln^2 3
  - \frac{17\sqrt{3}\pi^3}{72}
  - \frac{21}{2}\zeta_2
  + \frac{1}{3}\zeta_3
  \nonumber\\&&\mbox{}
  + 10\sqrt{3}{\rm Cl}_2\left(\frac{\pi}{3}\right)
  - 6\sqrt{3} {\rm Im}\left[ {\rm
     Li}_3\left(\frac{e^{-i\pi/6}}{\sqrt{3}}\right)\right] 
\,.
\label{eq:master2}
\end{eqnarray}
These expressions were used in~\cite{CheSte99} and coincide with 
the results listed in a recent work~\cite{FleKal99}.

As an outcome of the recurrence relations also products of one- and two-loop
integrals may appear. This makes in necessary to evaluate also the
${\cal O}(\varepsilon)$ part of the two-loop diagram shown in
Fig.~\ref{fig:master2}(a).
A straightforward Feynman parametrisation gives (in agreement
with~\cite{DavTau93}):
\begin{eqnarray}
T(M,M,M) &=&
\int
\frac{{\rm d}^Dp{\rm d}^Dk}{(M^2\pi)^D}
\frac{M^6}{
(p^2+M^2)
(k^2+M^2)
((p+k)^2+M^2)
}
\label{eq:master2l}
\\
&=&
  -\frac{3}{2\varepsilon^2}
  -\frac{9}{2\varepsilon}
  -\frac{21}{2}
  -\frac{3}{2}\zeta_2
  +2\sqrt{3}{\rm Cl}_2\left(\frac{\pi}{3}\right)
  +\varepsilon\left(
    - \frac{45}{2} 
    - \frac{\sqrt{3}\pi}{8} \ln^2 3
  \right.\nonumber\\&&\left.\mbox{}
    - \frac{35\sqrt{3}\pi^3}{216}
    - \frac{9}{2}\zeta_2
    + \zeta_3
    + 6\sqrt{3}{\rm Cl}_2\left(\frac{\pi}{3}\right)
    - 6\sqrt{3} {\rm Im}\left[ {\rm
        Li}_3\left(\frac{e^{-i\pi/6}}{\sqrt{3}}\right)\right] 
  \right)
\,.
\nonumber
\end{eqnarray}

All other integrals which have to solved are of one- and two-loop
tadpole type. The two-loop integrals are similar to the ones of
Eq.~(\ref{eq:master2l}), however, one or two lines are massless.
For these kind of integrals there exists a solution valid to any order
in $\varepsilon$ and for arbitrary
exponents of the propagators in the denominator.

A numerical check of the integrals in Eqs.~(\ref{eq:master1})
has been performed. A straightforward Feynman parameterization
leads to multidimensional integrals for which
a Monte Carlo integration routine~\cite{VEGAS}
has been used in order to obtain 5 significant digits.

As an independent check of the results for the diagrams pictured in
Fig.~\ref{fig:master2} the following method was used:
One of the massive lines is cut in such a way that the remaining
two-loop diagram has a threshold starting from $4M^2$.
Then the mass of the cut line is set to $M^\prime$ and an asymptotic expansion
for $M^\prime\ll M$ is performed.
The effective expansion parameter is $(M^\prime)^2/(4M^2)$.
Thus is very suggestive that even for $M^\prime=M$
a reasonably fast approximation to the exact result can be expected.
Actually the first 20 expansion terms for the diagrams in
Fig.~\ref{fig:master2} reproduce the first 10 digits
of the exact result.
Note that 
it only makes sense to apply this method if the diagram which results
from the original one after setting the mass of the selected line
to zero is computable analytically.


\section{Three-loop results}
\label{sec:3loop}

In this section the results of order $\alpha_s^3$ are discussed.
Altogether 131 diagrams contribute.
The evaluation in the limit $z\to0$ reduces to a naive Taylor expansion
in the external momentum. Thus one ends up with three-loop vacuum integrals
where the computation follows closely the lines of Section~\ref{sec:master}.
In the limit of large external momentum the rules of asymptotic expansion
(see, e.g., Refs.~\cite{Smi95,HarSte98} for reviews) have to be applied.
As a result for each diagram several subgraphs have to be considered
which leads to roughly 3500 diagrams to be evaluated.
It is clear that this cannot be done by hand and the use of
computer algebra becomes unavoidable.
The practical calculation has been performed with the package {\tt
  GEFICOM}~\cite{geficom}. It calls the programs {\tt LMP}~\cite{Har:diss},
respectively, {\tt EXP}~\cite{Sei:dipl} in order to perform the asymptotic
expansion.  
A recent summary of the available
software can be found in~\cite{HarSte98}.
In both limits we were able to evaluate the first seven terms in the
expansion, i.e. terms up to ${\cal O}(z^6)$ and ${\cal O}(1/z^6)$,
respectively, are available. Thereby Feynman gauge has been
adopted.

In principle the expansion terms of the colour structures 
$FFF$, $FFA$, $FFL$, $FFH$, $FAA$, $FAL$ und $FAH$ depend
on the QCD gauge parameter, $\xi$.
The remaining three structures contain two closed fermion loops
and thus $\xi$ drops out at an early stage of the calculation.
A straightforward computation of the $\xi$-dependent terms
produces huge expressions in intermediate steps and leads
quite fast to the limitations set by the soft- and hardware.
For the abelian case there exists a simple Ward identity
which allows a very fast calculation of the $\xi$-dependent 
part of the fermion propagator~\cite{LanLif}
\begin{eqnarray}
  \frac{d}{d \xi} 
  S_F^{0,QED} (q) &=& -e_0^2 \int \frac{d^D p}{i(2\pi)^D}
  \frac{1}{p^4} S_F^{0,QED} (q-p)
  \label{QED:WI}
  \,.
\end{eqnarray}
Within dimensional regularization Eq.~(\ref{QED:WI})
is an exact equation which relates the 
$\xi$-dependent part of the fermion propagator in   
$L+1$-loop order with the full propagator at $L$-loop order.
In fact, the calculation of the r.h.s. of (\ref{QED:WI})
for  general gauge is significantly faster than that of the very function 
$S_F^{QED}$ in (the simplest) Feynman gauge.
This is because of the ``factorizable'' nature of the integration with respect
to the momentum $p$. 

Using Eq.~(\ref{QED:WI}) we
have computed the ($\xi$-dependent) QED colour structures 
$FFF$, $FFL$ and $FFH$.
To our best knowledge there is not a simple generalization of
the identity~(\ref{QED:WI}) for QCD, i.e. the non-abelian contributions.
Thus the results for the corresponding colour factors can only be
analyzed for $\xi=0$.

Let us in a first step discuss the results for $\xi=0$ and afterwards
consider the QED-like diagrams for general $\xi$.

In Tabs.~\ref{tab:as3_1} and~\ref{tab:as3_2} the results for the 
individual colour structures can be found where $\xi=0$ has been adopted for 
the gauge parameter.
Only those results are listed which
include at least the terms of ${\cal O}(z^5)$ and ${\cal O}(1/z^4)$.
Furthermore we require that
the difference of the degree of the polynomial
in the numerator and denominator is less or equal to two.
The Pad\'e approximations which develop poles inside the unit plane
and where no appropriate cancellation takes place
are again represented by a dash.

{\footnotesize
\begin{table}[t]
\begin{center}
\begin{tabular}{|l|l||r|r|r|r|r|} 
\hline
$n$ & P.A.& $FFF$& $FFA$& $FFL$& $FFH$& $FAA$\\
\hline
$5$ & $[4/5]$ &  $2.0338$ &  $-2.0665$ &  $0.7347$ &  $2.4876$ &  $-16.2692$ \\
$5$ & $[4/6]$ &  $2.0272$ &  $-2.0031$ &  $0.7341$ &  $2.5147$ &  ${}^\star-16.4584$ \\
$5$ & $[5/4]$ &  $2.0310$ &  $-2.0632$ &  $0.7371$ &  $2.4752$ &  $-16.3826$ \\
$5$ & $[5/5]$ &  $2.0263$ &  $-1.9981$ &  $0.7343$ &  $2.5057$ &  $-16.4243$ \\
$5$ & $[5/6]$ &  $2.1191$ &  ${}^\star-1.9318$ &  ${}^\star0.7379$ &  $2.5305$ &  ${}^\star-16.4890$ \\
$5$ & $[6/4]$ &  $2.0344$ &  $-2.0025$ &  $0.7325$ &  $2.5623$ &  $-16.4381$ \\
$5$ & $[6/5]$ &  $2.1091$ &  ${}^\star-1.8018$ &  $0.7421$ &  $2.5281$ &  ${}^\star-16.4903$ \\
\hline 
$6$ & $[4/6]$ &  $2.0236$ &  ${}^\star-2.0587$ &  $0.7411$ &  $2.4954$ &  $-16.3748$ \\
$6$ & $[5/5]$ &  $2.0220$ &  $-2.0457$ &  $0.7423$ &  $2.4951$ &  $-16.3949$ \\
$6$ & $[5/6]$ &  ${}^\star1.9686$ &  ${}^\star-2.0304$ &  ${}^\star0.7372$ &  $2.5188$ &  ${}^\star-16.4554$ \\
$6$ & $[5/7]$ &  ${}^\star2.0648$ &  ${}^\star-2.0962$ &  $0.7601$ &  $1.1897$ & --- \\
$6$ & $[6/4]$ &  $2.0234$ &  $-1.9897$ &  $0.7410$ &  $2.4984$ &  $-16.4185$ \\
$6$ & $[6/5]$ &  $2.2334$ &  ${}^\star-2.0243$ &  ${}^\star0.7392$ &  $2.5182$ &  ${}^\star-16.4689$ \\
$6$ & $[6/6]$ &  ${}^\star2.0548$ &  ${}^\star-2.0858$ &  $0.7653$ &  ${}^\star2.6713$ &  ${}^\star-16.4411$ \\
$6$ & $[7/5]$ &  ${}^\star2.1493$ &  ${}^\star-2.1205$ &  $0.7567$ &  $2.3704$ &  $-16.3435$ \\
\hline 
\end{tabular}
\caption{\label{tab:as3_1}
Three-loop results for $z_m^X$ ($X=FFF,FFA,FFL,FFH,FAA$) for $\xi=0$.
The same notation as in Tab.~\ref{tab:12loop} has been adopted.
}
\end{center}
\end{table}
}

{\footnotesize
\begin{table}[t]
\begin{center}
\begin{tabular}{|l|l||r|r|r|r|r|} 
\hline
$n$ & P.A.& $FAL$& $FAH$& $FLL$& $FLH$& $FHH$\\
\hline
$5$ & $[4/5]$ & --- &  ${}^\star-1.4898$ &  ${}^\star-1.9890$ &  $-0.0630$ &  $-0.1806$ \\
$5$ & $[4/6]$ &  $13.4932$ &  $-1.3224$ & --- &  $-0.0708$ &  $-0.1958$ \\
$5$ & $[5/4]$ &  ${}^\star13.6711$ &  $-1.5551$ &  ${}^\star-1.9764$ &  $-0.0601$ &  $-0.1669$ \\
$5$ & $[5/5]$ &  $13.4519$ &  ${}^\star-1.5357$ & --- &  $-0.0686$ &  $-0.1948$ \\
$5$ & $[5/6]$ &  ${}^\star13.7944$ &  ${}^\star-1.5136$ &  ${}^\star-2.0122$ &  $-0.0703$ & --- \\
$5$ & $[6/4]$ &  $13.5259$ &  $-1.4147$ & --- &  $-0.0824$ &  $-0.3390$ \\
$5$ & $[6/5]$ &  ${}^\star13.6020$ &  ${}^\star-1.5133$ &  $-1.9646$ &  $-0.0690$ & --- \\
\hline 
$6$ & $[4/6]$ &  $13.3826$ &  ${}^\star-1.5250$ &  ${}^\star-1.9937$ &  $-0.0662$ &  $-0.1939$ \\
$6$ & $[5/5]$ &  $13.3533$ &  ${}^\star-1.5254$ &  ${}^\star-1.9985$ &  $-0.0660$ &  $-0.1930$ \\
$6$ & $[5/6]$ &  $12.9396$ &  ${}^\star-1.5183$ &  $-1.9767$ &  $-0.0695$ &  $-0.1928$ \\
$6$ & $[5/7]$ &  $13.3481$ & --- &  $-1.9696$ &  $-0.0486$ &  $-0.1887$ \\
$6$ & $[6/4]$ &  $13.4640$ &  $-1.4777$ &  ${}^\star-1.9981$ &  $-0.0676$ &  $-0.2154$ \\
$6$ & $[6/5]$ & --- &  ${}^\star-1.5173$ & --- &  $-0.0695$ &  $-0.1928$ \\
$6$ & $[6/6]$ &  $13.3138$ & --- &  $-1.9698$ &  $-0.0319$ &  $-0.1911$ \\
$6$ & $[7/5]$ &  $13.4209$ & --- &  $-1.9693$ &  $-0.0542$ &  $-0.1880$ \\
\hline 
\end{tabular}
\caption{\label{tab:as3_2}
Three-loop results for $z_m^X$ ($X=FAL,FAH,FLL,FLH,FHH$) for $\xi=0$.
The same notation as in Tab.~\ref{tab:12loop} has been adopted.
}
\end{center}
\end{table}
}

Very nice agreement between the different values in each column
of Tabs.~\ref{tab:as3_1} and ~\ref{tab:as3_2} can be found.
There is one obvious exception in the structure FFH namely 
the $[5/7]$ Pad\'e approximation where the sixth moment was used for the
construction. A closer look shows that there is a pole close to $z=1$
($\omega=1.105\ldots$, respectively, $z=0.997\ldots$) which explains the
deviation from the other results. 
In order to obtain the final numbers we will not take this value into
account.

From Tab.~\ref{tab:as3_2} one observes that the structures $FLH$ and $FHH$
are less stable than the others.
The reason for this is the same as
for $FH$ at two-loop order.
Again the threshold at $q^2=M^2$ is suppressed as compared to the one at
$q^2=9M^2$ which suggests that the expansion around $z=0$ should provide
a reasonable expansion parameter.
In Tab.~\ref{tab:as3_H}
the results for these two colour structures where no high energy terms were
used are shown. Actually almost
all Pad\'e approximants performed in the variable
$\omega$ develop poles for $|\omega|<1$. Thus we decided not to use them
for our analysis.
In analogy to the two-loop case the results are very stable
and only a small differences between the naive Taylor expansion
and diagonal Pad\'e results are observed.
Thus in the following the results of Tab.~\ref{tab:as3_H} are used
for the structures $FLH$ and $FHH$.
{\footnotesize
\begin{table}[t]
\begin{center}
\begin{tabular}{|l|l||r|r|r|r|} 
\hline
$n$ & P.A.& $ FFHz $& $ FAHz $& $ FLHz $& $ FHHz $\\
\hline
$4$ & $[1/3]$ &  $2.4135$ &  $-1.5047$ &  $-0.0669$ &  $-0.1922$ \\
$4$ & $[2/2]$ &  $2.4135$ &  ${}^\star-1.4790$ &  $-0.0670$ &  $-0.1922$ \\
$4$ & $[3/1]$ &  $2.4065$ &  $-1.5042$ &  $-0.0670$ &  $-0.1922$ \\
$4$ & $[4/0]$ &  $2.3531$ &  $-1.494$ &  $-0.0673$ &  $-0.1922$ \\
\hline 
$5$ & $[1/4]$ &  $2.4310$ &  $-1.5085$ &  $-0.0669$ &  $-0.1922$ \\
$5$ & $[2/3]$ &  ${}^\star2.4143$ &  $-1.5103$ &  $-0.0669$ &  $-0.1922$ \\
$5$ & $[3/2]$ &  $2.4364$ &  $-1.5102$ &  $-0.0668$ &  $-0.1922$ \\
$5$ & $[4/1]$ &  $2.4236$ &  $-1.5080$ &  $-0.0669$ &  $-0.1922$ \\
$5$ & $[5/0]$ &  $2.3765$ &  $-1.4990$ &  $-0.0671$ &  $-0.1922$ \\
\hline 
$6$ & $[1/5]$ &  $2.4410$ &  $-1.5109$ &  $-0.0668$ &  $-0.1922$ \\
$6$ & $[2/4]$ &  $2.4473$ &  $-1.5126$ &  ${}^\star-0.0669$ &  $-0.1922$ \\
$6$ & $[3/3]$ &  $2.4488$ &  $-1.5120$ &  $-0.0668$ &  $-0.1922$ \\
$6$ & $[4/2]$ &  $2.4474$ &  $-1.5126$ &  $-0.0668$ &  $-0.1922$ \\
$6$ & $[5/1]$ &  $2.4352$ &  $-1.5105$ &  $-0.0668$ &  $-0.1922$ \\
$6$ & $[6/0]$ &  $2.3933$ &  $-1.5024$ &  $-0.0669$ &  $-0.1922$ \\
\hline 
\end{tabular}
\caption{\label{tab:as3_H}
  Results for the colour structures involving a closed heavy fermion loop.
  Only the moments for $z\to0$
  have been implemented into the Pad\'e analysis.
}
\end{center}
\end{table}
}

It is tempting to consider also for the colour factors $C_F^2T$ and
$C_FC_AT$ the results which exclusively contain the information 
from the expansion around $z\to0$. They are also displayed in
Tab.~\ref{tab:as3_H}. There is perfect agreement with the numbers
extracted from Tabs.~\ref{tab:as3_1} and~\ref{tab:as3_2},
the errors, however, are significantly smaller.
Thus we decided to take also for these colour structures the results of
Tab.~\ref{tab:as3_H}.

Finally we get the following results if $\xi=0$ is adopted:
\begin{eqnarray}
\renewcommand{\arraystretch}{1.2}
\begin{array}{lll}
\displaystyle
z_m^{FFF} \,\,=\,\, 2.1(3)
\,,
&
\displaystyle
z_m^{FFA} \,\,=\,\, -2.0(4)
\,,
&
\displaystyle
z_m^{FFL} \,\,=\,\, 0.74(4)
\,,
\\
\displaystyle
z_m^{FFH} \,\,=\,\, 2.4(1)
\,,
&
z_m^{FAA} \,\,=\,\, -16.4(3)
\,,
&
z_m^{FAL} \,\,=\,\, 13(1)
\,,
\\
z_m^{FAH} \,\,=\,\, -1.51(5)
\,,
&
z_m^{FLL} \,\,=\,\, -1.98(6)
\,,
&
z_m^{FLH} \,\,=\,\, -0.0669(6)
\,,
\\
z_m^{FHH} \,\,=\,\, -0.1922(0)
\,,
\label{eq:zm3loop}
\end{array}
\end{eqnarray}
where the error has again been obtained by doubling the spread of the Pad\'e
approximants.

Let us now analyze the QED-like diagrams, i.e. the structures
$FFF$, $FFL$ and $FFH$ for arbitrary gauge parameter.
In analogy to the one- and two-loop case considered in
Section~\ref{sec:12loop}
we vary $\xi$ between $-5$ and $+5$. The corresponding results can be found
in Tabs.~\ref{tab:QEDxi_1} and~\ref{tab:QEDxi_2}.

\begin{table}[t]
\begin{center}
\begin{tabular}{cc}
{\tiny
\begin{tabular}{|l|l||r|r|r|} 
\hline
$n$ & P.A.& $ FFF $& $ FFL $& $ FFH $\\
\hline
$5$ & $[4/5]$ &  $2.1093$ &  $1.0402$ &  $2.5059$ \\
$5$ & $[4/6]$ &  $2.0926$ &  $1.0354$ &  $2.4652$ \\
$5$ & $[5/4]$ &  $2.1082$ &  $1.0430$ &  $2.5013$ \\
$5$ & $[5/5]$ &  $2.0923$ &  ${}^\star1.0443$ &  $2.4650$ \\
$5$ & $[5/6]$ &  ${}^\star1.5554$ &  $1.0229$ &  $3.1456$ \\
$5$ & $[6/4]$ &  $2.0954$ &  $1.0193$ &  $2.4717$ \\
$5$ & $[6/5]$ &  $2.6737$ &  $1.0230$ &  $2.8145$ \\
\hline 
$6$ & $[4/6]$ &  $2.0985$ &  ${}^\star1.0540$ &  $2.4565$ \\
$6$ & $[5/5]$ &  $2.1016$ &  ${}^\star1.0602$ &  $2.4248$ \\
$6$ & $[5/6]$ &  ${}^\star2.0780$ &  $1.0016$ &  ${}^\star2.4470$ \\
$6$ & $[5/7]$ &  ${}^\star2.1475$ &  ${}^\star1.0204$ &  ${}^\star2.5472$ \\
$6$ & $[6/4]$ &  $2.1000$ &  $1.0283$ &  $2.4623$ \\
$6$ & $[6/5]$ &  ${}^\star2.0498$ &  $1.0103$ &  ${}^\star2.3488$ \\
$6$ & $[6/6]$ &  ${}^\star2.1400$ &  ${}^\star1.0278$ &  ${}^\star2.5325$ \\
$6$ & $[7/5]$ &  ${}^\star2.1828$ &  ${}^\star1.0118$ &  ${}^\star2.6241$ \\
\hline 
\end{tabular}
}
&
{\tiny
\begin{tabular}{|l|l||r|r|r|} 
\hline
$n$ & P.A.& $ FFF $& $ FFL $& $ FFH $\\
\hline
$5$ & $[4/5]$ &  $2.0599$ &  $0.8596$ &  $2.4949$ \\
$5$ & $[4/6]$ &  $2.0447$ &  $0.8527$ &  $2.4978$ \\
$5$ & $[5/4]$ &  $2.0580$ &  $0.8599$ &  $2.4865$ \\
$5$ & $[5/5]$ &  $2.0444$ &  $0.8549$ &  $2.4932$ \\
$5$ & $[5/6]$ &  $2.3098$ &  $0.8545$ &  $2.5776$ \\
$5$ & $[6/4]$ &  $2.0489$ &  $0.8468$ &  $2.5242$ \\
$5$ & $[6/5]$ &  $2.2055$ &  $0.8558$ &  $2.5769$ \\
\hline 
$6$ & $[4/6]$ &  $2.0434$ & --- &  $2.4867$ \\
$6$ & $[5/5]$ &  $2.0369$ &  ${}^\star0.8575$ &  $2.4863$ \\
$6$ & $[5/6]$ &  ${}^\star2.0280$ &  $0.8534$ &  $2.5959$ \\
$6$ & $[5/7]$ &  ${}^\star2.0854$ &  $0.8568$ &  ${}^\star2.5867$ \\
$6$ & $[6/4]$ &  $2.0441$ &  $0.8537$ &  $2.4863$ \\
$6$ & $[6/5]$ &  ${}^\star1.9423$ &  $0.8535$ &  $2.5567$ \\
$6$ & $[6/6]$ &  ${}^\star2.0777$ &  $0.8570$ &  ${}^\star2.5503$ \\
$6$ & $[7/5]$ &  ${}^\star2.1340$ &  $0.8569$ &  ${}^\star3.5102$ \\
\hline 
\end{tabular}
}
\end{tabular}
\caption{\label{tab:QEDxi_1}
  Results for the structures $FFF$, $FFL$ and $FFH$
  where for the gauge parameter
  the values $\xi=-5$ (left table) and $\xi=-2$ (right table) have been
  adopted. 
}
\end{center}
\end{table}
\begin{table}[t]
\begin{center}
\begin{tabular}{cc}
{\tiny
\begin{tabular}{|l|l||r|r|r|} 
\hline
$n$ & P.A.& $ FFF $& $ FFL $& $ FFH $\\
\hline
$5$ & $[4/5]$ &  $1.9184$ &  $0.6080$ &  $2.4799$ \\
$5$ & $[4/6]$ &  $1.8976$ &  $0.6182$ &  $2.5284$ \\
$5$ & $[5/4]$ &  $1.9030$ &  $0.6138$ &  $2.4614$ \\
$5$ & $[5/5]$ &  $1.8970$ &  $0.6187$ &  $2.5146$ \\
$5$ & $[5/6]$ &  $1.9309$ &  ${}^\star0.6178$ &  $2.5105$ \\
$5$ & $[6/4]$ &  $1.9061$ &  $0.6187$ &  $2.6074$ \\
$5$ & $[6/5]$ &  $1.9309$ &  ${}^\star0.6267$ &  $2.5030$ \\
\hline 
$6$ & $[4/6]$ &  $1.8906$ &  $0.6384$ &  $2.5005$ \\
$6$ & $[5/5]$ &  $1.8878$ &  $0.6557$ &  $2.4993$ \\
$6$ & $[5/6]$ & --- &  ${}^\star0.6091$ &  $2.5076$ \\
$6$ & $[5/7]$ &  ${}^\star1.9425$ &  $0.6720$ &  $2.4446$ \\
$6$ & $[6/4]$ &  $1.8897$ &  $0.6311$ &  $2.5098$ \\
$6$ & $[6/5]$ &  $1.9346$ &  ${}^\star0.6175$ &  $2.5075$ \\
$6$ & $[6/6]$ &  ${}^\star1.9248$ &  $0.6886$ &  $2.4268$ \\
$6$ & $[7/5]$ &  ${}^\star3.0282$ &  $0.6587$ &  $2.4514$ \\
\hline 
\end{tabular}
}
&
{\tiny
\begin{tabular}{|l|l||r|r|r|} 
\hline
$n$ & P.A.& $ FFF $& $ FFL $& $ FFH $\\
\hline
$5$ & $[4/5]$ &  $0.9431$ &  $0.4164$ &  $2.4671$ \\
$5$ & $[4/6]$ &  $1.2669$ &  $0.4493$ &  $2.5427$ \\
$5$ & $[5/4]$ &  $1.3734$ &  $0.4280$ &  $2.4317$ \\
$5$ & $[5/5]$ &  $1.2631$ &  $0.4558$ &  $2.5229$ \\
$5$ & $[5/6]$ &  $1.2435$ &  ${}^\star0.4313$ &  $2.4983$ \\
$5$ & $[6/4]$ &  $1.2642$ &  $0.4492$ &  $2.7025$ \\
$5$ & $[6/5]$ &  $1.2325$ &  ${}^\star0.4503$ &  $2.4846$ \\
\hline 
$6$ & $[4/6]$ &  $1.2467$ &  $0.5049$ &  $2.5046$ \\
$6$ & $[5/5]$ &  $1.2236$ &  $0.6468$ &  $2.5022$ \\
$6$ & $[5/6]$ &  $1.2332$ &  ${}^\star0.3972$ &  $2.5029$ \\
$6$ & $[5/7]$ &  $1.1881$ &  $0.5539$ &  $2.4788$ \\
$6$ & $[6/4]$ &  $1.2428$ &  $0.4712$ &  $2.5305$ \\
$6$ & $[6/5]$ &  $1.2510$ &  ${}^\star0.4210$ &  $2.5020$ \\
$6$ & $[6/6]$ &  $1.1799$ &  $0.5975$ &  $2.4780$ \\
$6$ & $[7/5]$ &  $1.1955$ &  $0.5159$ &  $2.4782$ \\
\hline 
\end{tabular}
}
%
\end{tabular}
\caption{\label{tab:QEDxi_2}
  Results for the structures $FFF$, $FFL$ and $FFH$
  where for the gauge parameter
  the values $\xi=+2$ (left table) and $\xi=+5$ (right table) have been
  adopted. 
}
\end{center}
\end{table}

In the case $FFH$ we actually prefer to use only the low-energy moments.
The corresponding results for $\xi=-5,-2,0,+2$ and $+5$ can be found in
Tab.~\ref{tab:FFHz_xi}. For all choices very stable results with a spread
below 3\% are obtained. Furthermore they are all consistent with each other.
{\footnotesize
\begin{table}[ht]
\begin{center}
\begin{tabular}{|l|l||r|r|r|r|} 
\hline
$n$ & P.A.& $\xi=-5$ & $\xi=-2$ & $\xi=+2$ & $\xi=+5$ \\
\hline
$4$ & $[1/3]$ &  $2.4174$ &  $2.4145$ &  $2.4135$ &  $2.4161$ \\
$4$ & $[2/2]$ &  $2.4182$ &  $2.4146$ &  $2.4135$ &  $2.4162$ \\
$4$ & $[3/1]$ &  $2.4165$ &  $2.4102$ &  $2.4032$ &  $2.3985$ \\
$4$ & $[4/0]$ &  $2.3648$ &  $2.3578$ &  $2.3484$ &  $2.3414$ \\
\hline 
$5$ & $[1/4]$ &  $2.4336$ &  $2.4317$ &  $2.4306$ &  $2.4304$ \\
$5$ & $[2/3]$ &  $2.4271$ &  ${}^\star2.4204$ &  ${}^\star2.4029$ & ---\\
$5$ & $[3/2]$ &  $2.4436$ &  $2.4386$ &  $2.4345$ &  $2.4319$ \\
$5$ & $[4/1]$ &  $2.4298$ &  $2.4259$ &  $2.4215$ &  $2.4184$ \\
$5$ & $[5/0]$ &  $2.3847$ &  $2.3798$ &  $2.3733$ &  $2.3684$ \\
\hline 
$6$ & $[1/5]$ &  $2.4423$ &  $2.4413$ &  $2.4408$ &  $2.4406$ \\
$6$ & $[2/4]$ &  $2.4478$ &  $2.4474$ &  $2.4478$ &  $2.4525$ \\
$6$ & $[3/3]$ &  $2.4525$ &  $2.4497$ &  $2.4485$ &  $2.4526$ \\
$6$ & $[4/2]$ &  $2.4523$ &  $2.4490$ &  $2.4460$ &  $2.4441$ \\
$6$ & $[5/1]$ &  $2.4394$ &  $2.4368$ &  $2.4337$ &  $2.4315$ \\
$6$ & $[6/0]$ &  $2.3992$ &  $2.3957$ &  $2.3909$ &  $2.3873$ \\
\hline 
\end{tabular}
\caption{\label{tab:FFHz_xi}
  Results for $FFH$ where only the low-energy moments are taken into account.
  Different choices for the gauge parameter are adopted.
}
\end{center}
\end{table}
}

\begin{figure}[t]
  \begin{center}
    \begin{tabular}{c}
      \leavevmode
      \epsfxsize=14.cm
      \epsffile[90 420 510 730]{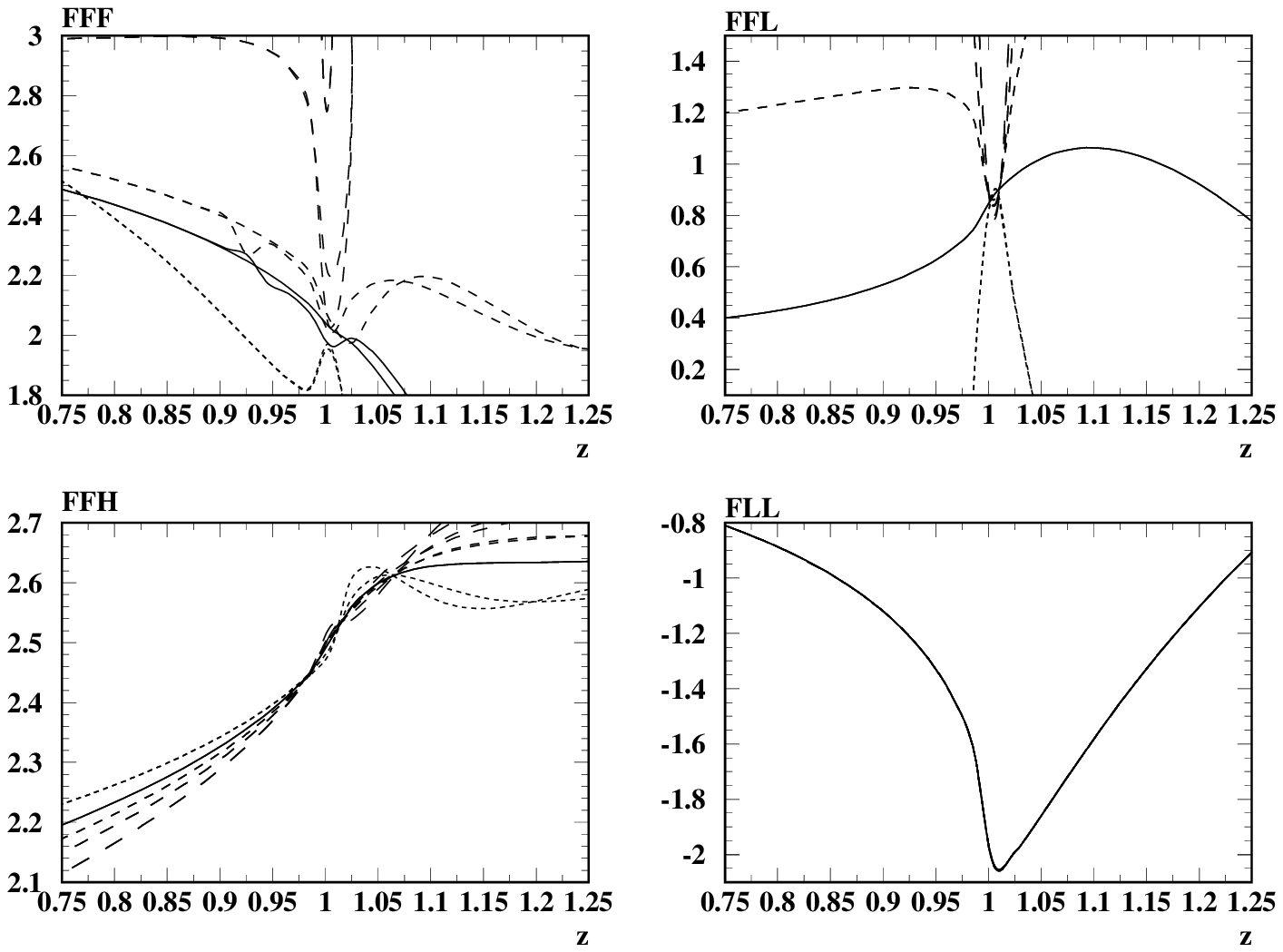}
    \end{tabular}
  \end{center}
  \caption{\label{fig:QEDxidep}
    $z$ dependence of $-g(z)$ for the structures $FFF$, $FFL$ and $FFH$ for
    different values of $\xi$. The solid curves belong to $\xi=-2$.
    The dashed curves correspond to $\xi=-5,0,+2,+5$ (from short to long
    dashes).
    }
\end{figure}

It is instructive to examine also at three-loop order the $z$ dependence
around $z=1$ for different values of $\xi$ which is done in
Fig.~\ref{fig:QEDxidep} for the $\xi$-dependent colour
structures $FFF$, $FFL$ and $FFH$.
The result for $FLL$ which does not depend on $\xi$
is plotted for comparison.
Actually the corresponding part of the 
function $-g(z)$ is plotted in the vicinity of $z=1$.
The ten curves in each plot correspond to $\xi=-5,-2,0,+2$ and $\xi=+5$
where for each value of $\xi$ two different Pad\'e results are shown.

It can be seen that 
as for the order $\alpha_s$ and $\alpha_s^2$ cases also here the choice
$\xi=-2$ (solid curve) exhibits the smoothest behaviour for $z=1$
which was already expected from the stable behaviour of the numbers in
Tabs.~\ref{tab:QEDxi_1} and~\ref{tab:QEDxi_2}.
Thus it can be expected that for this values our procedure works best.
However, we want to stress that --- in analogy to the lower order analysis ---
the value
$\xi=0$ looks very promising and also for this choice reasonable results can
be expected.

Adopting the choice $\xi=-2$ we extract from Tabs.~\ref{tab:QEDxi_1}
and~\ref{tab:FFHz_xi} the following results:
\begin{eqnarray}
  z_m^{FFF} = 2.1(4)\,,\qquad
  z_m^{FFL} = 0.8(2)\,,\qquad
  z_m^{FFH} = 2.4(1)\,.
  \label{eq:resQED3loop}
\end{eqnarray}
Note that the colour structure $FFL$ demonstrates a rather strong
dependence on the gauge parameter. To take this into account we assigned 
somewhat extended errors to this quantity  in  Eq.~(\ref{eq:resQED3loop}).

Note that the results of~(\ref{eq:zm3loop}) and ~(\ref{eq:resQED3loop})
are valid for $\mu^2=M^2$. 
With the help of the
two-loop $\beta$~\cite{2lbeta}
and three-loop $\gamma_m$~\cite{3lgm} function it is possible to restore
the $\ln(\mu^2/M^2)$ terms at order $\alpha_s^3$. Thus Eq.~(\ref{eq:mmsos})
reads
\begin{eqnarray}
  \frac{m(\mu)}{M} &=&
  1 
  + \frac{\alpha_s^{(n_f)}(\mu)}{\pi} z_m^{(1)}(\mu)
  + \left(\frac{\alpha_s^{(n_f)}(\mu)}{\pi}\right)^2 z_m^{(2)}(\mu)
  + \left(\frac{\alpha_s^{(n_f)}(\mu)}{\pi}\right)^3
  \Bigg\{ z_m^{(3)}(M) +
  \nonumber\\&&\mbox{}
  C_F^3\left[
    - \frac{9}{128}\lmM^3
    - \frac{27}{128}\lmM^2
    + \left(
      -\frac{489}{512}
      + \frac{45}{32}\zeta_2
      - \frac{9}{4}\zeta_2\ln2
      + \frac{9}{16}\zeta_3
      \right)\lmM
  \right]
  \nonumber\\&&\mbox{}
  + C_F^2C_A\left[
    \frac{33}{128}\lmM^3
    +\frac{109}{64}\lmM^2
    +\left(
      \frac{5813}{1536} 
      - \frac{61}{16}\zeta_2 
      + \frac{53}{8}\zeta_2\ln2
      - \frac{53}{32}\zeta_3
    \right)\lmM
  \right]
  \nonumber\\&&\mbox{}
  +C_F^2Tn_l\left[
    - \frac{3}{32}\lmM^3
    - \frac{13}{32}\lmM^2
    + \left(
      \frac{65}{384}
      +\frac{7}{8}\zeta_2
      - 2\zeta_2\ln2
      - \frac{1}{4}\zeta_3
    \right)\lmM
  \right]
  \nonumber\\&&\mbox{}
  +C_F^2T\left[
    - \frac{3}{32}\lmM^3
    - \frac{13}{32}\lmM^2
    + \left(
      -\frac{151}{384} 
      + 2\zeta_2 
      - 2\zeta_2 \ln2
      - \frac{1}{4}\zeta_3
    \right)\lmM
  \right]
  \nonumber\\&&\mbox{}
  +C_FC_A^2\left[
    - \frac{121}{576}\lmM^3
    - \frac{2341}{1152}\lmM^2
    + \left(
      - \frac{13243}{1728} 
      + \frac{11}{12}\zeta_2 
      - \frac{11}{4}\zeta_2\ln2
      + \frac{11}{16}\zeta_3
    \right)\lmM
  \right]
  \nonumber\\&&\mbox{}
  +C_FC_ATn_l\left[
    \frac{11}{72}\lmM^3
    +\frac{373}{288}\lmM^2
    +\left(
      \frac{869}{216}
      + \frac{7}{12}\zeta_2
      + \zeta_2\ln2
      + \frac{1}{2}\zeta_3
    \right)\lmM
  \right]
  \nonumber\\&&\mbox{}
  +C_FC_AT\left[
    + \frac{11}{72}\lmM^3
    + \frac{373}{288}\lmM^2
    + \left(
      \frac{583}{108}
      - \frac{13}{6}\zeta_2
      + \zeta_2\ln2
      + \frac{1}{2}\zeta_3
    \right) \lmM
  \right]
  \nonumber\\&&\mbox{}
  +C_FT^2n_l^2\left[
    - \frac{1}{36}\lmM^3
    - \frac{13}{72}\lmM^2
    + \left(
      -\frac{89}{216}
      -\frac{1}{3}\zeta_2
    \right)\lmM
  \right]
  \nonumber\\&&\mbox{}
  +C_FT^2n_l\left[
    - \frac{1}{18}\lmM^3
    - \frac{13}{36}\lmM^2
    + \left(
      -\frac{143}{108}
      + \frac{1}{3}\zeta_2
    \right)\lmM
  \right]
  \nonumber\\&&\mbox{}
  +C_FT^2\left[
    - \frac{1}{36}\lmM^3
    - \frac{13}{72}\lmM^2
    + \left(
      -\frac{197}{216}
      +\frac{2}{3}\zeta_2
    \right)\lmM
  \right]
  \Bigg\}
  \,,
  \label{eq:zmlog}
\end{eqnarray}
with $\lmM=\ln\mu^2/M^2$.
$n_f=n_l+1$ is the number of active quark flavours.
Very often the scale-invariant mass, $\mu_m$ defined through 
$\mu_m=m(\mu_m)$ is of interest. Thus iterating~(\ref{eq:zmlog}) leads to
\begin{eqnarray}
  \frac{\mu_m}{M} &=&
  1 
  + \frac{\alpha_s^{(n_f)}(M)}{\pi} z_m^{(1)}(M)
  + \left(\frac{\alpha_s^{(n_f)}(M)}{\pi}\right)^2\left[
    z_m^{(2)}(M)
    + C_F^2 \frac{3}{2}
  \right]
  + \left(\frac{\alpha_s^{(n_f)}(M)}{\pi}\right)^3
  \nonumber\\&&\mbox{}
  \Bigg[ z_m^{(3)}(M)
  +C_F^3\left(
    - \frac{741}{256} 
    + \left(\frac{45}{16} - \frac{9}{2}\ln2\right)\zeta_2
    + \frac{9}{8}\zeta_3
  \right)
  +C_F^2C_A\left(
    \frac{4885}{768} - \left(\frac{3}{4} 
    \right.\right.\nonumber\\&&\left.\left.\mbox{}
      - \frac{9}{4}\ln2\right)\zeta_2
    - \frac{9}{16}\zeta_3
  \right)
  +C_F^2T\left(
    -\frac{509}{192} + \frac{3}{2}\zeta_2
  \right)
  +C_F^2Tn_l\left(
    -\frac{293}{192} - \frac{3}{4}\zeta_2
  \right)
  \Bigg]
  \,.
  \label{eq:zmzm}
\end{eqnarray}
Inverting Eq.~(\ref{eq:zmlog}) leads to
\begin{eqnarray}
  \frac{M}{m(\mu)} &=&
  1 
  + \frac{\alpha_s^{(n_f)}(\mu)}{\pi} \left(
    -z_m^{(1)}(M) + \frac{3}{4}\lmm
  \right)
  + \left(\frac{\alpha_s^{(n_f)}(\mu)}{\pi}\right)^2\Bigg\{
  -z_m^{(2)}(M)
  \nonumber\\&&\mbox{}
  +C_F^2\left[
    - \frac{1}{2}
    -\frac{9}{32}\lmm
    +\frac{9}{32}\lmm^2
  \right]
  +C_FC_A\left[
    \frac{185}{96}\lmm
    +\frac{11}{32}\lmm^2
  \right]
  \nonumber\\&&\mbox{}
  +C_FT\left(n_l+1\right)\left[
    -\frac{13}{24}\lmm
    -\frac{1}{8}\lmm^2
  \right]
  \Bigg\}
  \nonumber\\&&\mbox{}
  + \left(\frac{\alpha_s^{(n_f)}(\mu)}{\pi}\right)^3
  \Bigg\{ -z_m^{(3)}(M)
  +C_F^3\left[
    \frac{201}{256} 
    + \left(\frac{15}{16}\zeta_2
      - \frac{3}{2}\ln2
    \right)\zeta_2
    + \frac{3}{8}\zeta_3
    + \left(
      \frac{495}{512} 
    \right.\right.\nonumber\\&&\left.\left.\mbox{}
      + \left(\frac{45}{32}
        - \frac{9}{4}\ln2
      \right)\zeta_2
      + \frac{9}{16}\zeta_3
    \right)\lmm
    - \frac{63}{128} \lmm^2
    + \frac{9}{128} \lmm^3
  \right]
  +C_F^2C_A\left[
    -\frac{1849}{768} 
    - \left(
      \frac{1}{4} 
    \right.\right.\nonumber\\&&\left.\left.\mbox{}
      - \frac{3}{4}\ln2
    \right)\zeta_2
    - \frac{3}{16}\zeta_3
    + \left(
      -\frac{4219}{1536} 
      + \left(\frac{49}{16}
        - \frac{35}{8}\ln2
      \right)\zeta_2
      + \frac{35}{32}\zeta_3
    \right)\lmm
    + \frac{21}{64}\lmm^2
    \right.\nonumber\\&&\left.\mbox{}
    + \frac{33}{128}\lmm^3
  \right]
  +C_F^2Tn_l\left[
    \frac{137}{192} 
    - \frac{1}{4}\zeta_2
    + \left(
      -\frac{91}{384} 
      +\zeta_2\left(-\frac{13}{8}
        + 2\ln2
      \right)
      + \frac{1}{4}\zeta_3
    \right)\lmm
  \right.\nonumber\\&&\left.\mbox{}
    - \frac{3}{32}\lmm^2
    - \frac{3}{32}\lmm^3
  \right]
  +C_F^2T\left[
    \frac{65}{192} 
    + \frac{1}{2}\zeta_2
    + \left(
      -\frac{307}{384} 
      +\zeta_2\left(-\frac{1}{2}
        + 2\ln2
      \right)
    \right.\right.\nonumber\\&&\left.\left.\mbox{}
      + \frac{1}{4}\zeta_3
    \right)\lmm
    - \frac{3}{32}\lmm^2
    - \frac{3}{32}\lmm^3
  \right]
  +C_FC_A^2\left[
    + \left(
      \frac{13243}{1728} 
      +\left(-\frac{11}{12}
        + \frac{11}{4}\ln2
      \right)\zeta_2
    \right.\right.\nonumber\\&&\left.\left.\mbox{}
      - \frac{11}{16}\zeta_3
    \right)\lmm
    + \frac{2341}{1152}\lmm^2
    + \frac{121}{576}\lmm^3
  \right]
  \nonumber\\&&\mbox{}
  +C_FC_ATn_l\left[
    \left(
      -\frac{869}{216} 
      -\left(\frac{7}{12}
        + \ln2
      \right)\zeta_2
      - \frac{1}{2}\zeta_3
    \right)\lmm
    - \frac{373}{288}\lmm^2
    - \frac{11}{72}\lmm^3
  \right]
  \nonumber\\&&\mbox{}
  +C_FC_AT\left[
    \left(
      -\frac{583}{108} 
      +\left(\frac{13}{6}
        - \ln2
      \right)\zeta_2
      - \frac{1}{2}\zeta_3
    \right)\lmm
    - \frac{373}{288}\lmm^2
    - \frac{11}{72}\lmm^3
  \right]
  \nonumber\\&&\mbox{}
  +C_FT^2n_l^2\left[
    \left(
      \frac{89}{216} 
      + \frac{1}{3}\zeta_2
    \right)\lmm
    + \frac{13}{72}\lmm^2
    + \frac{1}{36}\lmm^3
  \right]
  \nonumber\\&&\mbox{}
  +C_FT^2n_l\left[
    \left(
      \frac{143}{108} 
      - \frac{1}{3}\zeta_2
    \right)\lmm
    + \frac{13}{36}\lmm^2
    + \frac{1}{18}\lmm^3
  \right]
  \nonumber\\&&\mbox{}
  +C_FT^2\left[
    \left(
      \frac{197}{216} 
      - \frac{2}{3}\zeta_2
    \right)\lmm
    + \frac{13}{72}\lmm^2
    + \frac{1}{36}\lmm^3
  \right]
\Bigg\}
  \,,
  \label{eq:zminv}
\end{eqnarray}
with $\lmm=\ln\mu^2/m^2(\mu)$.

If one is interested in the Eqs.~(\ref{eq:zmlog}),~(\ref{eq:zmzm})
or~(\ref{eq:zminv}) the results of~(\ref{eq:zm3loop})
and~(\ref{eq:resQED3loop}) can be used.
However, for most practical applications only the values $N_c=3$,
i.e. $C_F=4/3$ and $C_A=3$, and $T=1/2$ are of interest.
Simply adding the results of the individual colour factors
would lead to a big overestimation of the error.
A closer look to the numbers in Eqs.~(\ref{eq:zm3loop})
and~(\ref{eq:resQED3loop}) also
indicates that once the 
numerical values for the colour factors and the number of light quarks
are inserted
numerical cancellations might occur which result in a loss of accuracy.
Thus it is much more
promising to add in a first step the results for the moments,
of course, taking into account the proper colour factor.
Afterwards the Pad\'e approximations are computed.
As the contributions with and without heavy quark loops
are treated differently
we consider the sum of both sets separately as a function of $n_l$.
In Tab.~\ref{tab:TOTL} the results for the sum of the structures
$FFF$, $FFA$, $FFL$, $FAA$, $FAL$ and $FLL$ is shown where $n_l$ is varied
from 0 to 5. 
The analog combination of the remaining contributions ($FFH$, $FAH$, $FLH$,
$FHH$) can be found in Tab.~\ref{tab:TOTH}.
{\footnotesize
\begin{table}[t]
\begin{center}
\begin{tabular}{|l|l||r|r|r|r|r|r|} 
\hline
$n$ & P.A.& $ n_l=0 $& $ n_l=1 $& $ n_l=2 $& $ n_l=3 $& $ n_l=4 $& $ n_l=5 $\\
\hline
$5$ & $[4/5]$ &  $-200.2787$ &  $-173.6663$ &  $-148.3787$ &  $-124.4156$ &  $-101.7771$ &  $-80.4628$ \\
$5$ & $[4/6]$ &  $-201.6419$ &  $-174.8844$ &  $-149.4553$ &  $-125.3553$ &  $-102.5862$ &  $-81.1625$ \\
$5$ & $[5/4]$ &  $-203.9394$ &  $-176.7290$ &  $-150.8970$ &  $-126.4411$ &  $-103.3591$ &  $-81.6482$ \\
$5$ & $[5/5]$ &  $-201.4721$ &  $-174.7445$ &  $-149.3422$ &  $-125.2644$ &  $-102.5104$ &  $-81.0786$ \\
$5$ & $[5/6]$ &  $-198.7799$ &  $-172.8884$ &  $-148.2001$ & --- &  ${}^\star-102.2739$ &  ${}^\star-81.0336$ \\
$5$ & $[6/4]$ &  $-202.8435$ &  $-175.8651$ &  $-150.2298$ &  $-125.9387$ &  $-102.9929$ &  $-81.3939$ \\
\hline 
$6$ & $[4/6]$ &  $-201.0906$ &  $-174.3880$ &  $-149.0165$ &  $-124.9749$ &  $-102.2619$ &  $-80.8758$ \\
$6$ & $[5/5]$ &  $-200.9265$ &  $-174.2458$ &  $-148.8927$ &  $-124.8668$ &  $-102.1673$ &  $-80.7929$ \\
$6$ & $[5/6]$ &  ${}^\star-200.4927$ &  ${}^\star-173.9600$ &  ${}^\star-148.7433$ &  ${}^\star-124.8358$ &  ${}^\star-102.2290$ &  ${}^\star-80.9131$ \\
$6$ & $[5/7]$ &  $-200.3603$ &  $-173.7018$ &  $-148.3764$ &  $-124.3940$ &  $-101.7753$ &  $-80.5533$ \\
$6$ & $[6/4]$ &  $-201.6970$ &  $-174.9293$ &  $-149.4861$ &  $-125.3673$ &  $-102.5725$ &  $-81.1016$ \\
$6$ & $[6/6]$ &  $-200.3195$ &  $-173.6857$ &  $-148.3751$ &  $-124.3879$ &  $-101.7244$ &  $-80.3848$ \\
$6$ & $[7/5]$ &  $-202.1300$ &  $-175.2569$ &  $-149.7173$ &  $-125.5125$ &  $-102.6443$ &  $-81.1143$ \\
\hline 
\end{tabular}
\caption{\label{tab:TOTL}
  Pad\'e results for the sum of those contributions which don't have a closed
  heavy fermion loop. $n_l$ has been varied from 0 to 5.
}
\end{center}
\end{table}
}
{\footnotesize
\begin{table}[t]
\begin{center}
\begin{tabular}{|l|l||r|r|r|r|r|r|} 
\hline
$n$ & P.A.&$n_l=0 $& $n_l=1  $& $n_l=2  $& $n_l=3  $& $n_l=4  $& $n_l=5  $\\
\hline
$4$ & $[1/3]$ &  $-0.9345$ &  $-0.9572$ &  $-0.9798$ &  $-1.0024$ &  $-1.0249$ &  $-1.0475$ \\
$4$ & $[2/2]$ &  $-0.9321$ &  $-0.9546$ &  $-0.9770$ &  $-0.9995$ &  $-1.0218$ &  $-1.0442$ \\
$4$ & $[3/1]$ &  $-0.9324$ &  $-0.9551$ &  $-0.9777$ &  $-1.0003$ &  $-1.0229$ &  $-1.0455$ \\
$4$ & $[4/0]$ &  $-0.9604$ &  $-0.9828$ &  $-1.0053$ &  $-1.0277$ &  $-1.0501$ &  $-1.0725$ \\
\hline 
$5$ & $[1/4]$ &  $-0.9271$ &  $-0.9495$ &  $-0.9720$ &  $-0.9944$ &  $-1.0169$ &  $-1.0393$ \\
$5$ & $[2/3]$ &  $-0.9219$ &  $-0.9440$ &  $-0.9661$ &  $-0.9882$ &  $-1.0103$ &  $-1.0324$ \\
$5$ & $[3/2]$ &  $-0.9086$ &  $-0.9347$ &  $-0.9591$ &  $-0.9827$ &  $-1.0060$ &  $-1.0290$ \\
$5$ & $[4/1]$ &  $-0.9254$ &  $-0.9478$ &  $-0.9703$ &  $-0.9927$ &  $-1.0151$ &  $-1.0375$ \\
$5$ & $[5/0]$ &  $-0.9495$ &  $-0.9719$ &  $-0.9942$ &  $-1.0166$ &  $-1.0389$ &  $-1.0613$ \\
\hline 
$6$ & $[1/5]$ &  $-0.9217$ &  $-0.9441$ &  $-0.9665$ &  $-0.9888$ &  $-1.0112$ &  $-1.0336$ \\
$6$ & $[2/4]$ &  $-0.9140$ &  $-0.9364$ &  $-0.9589$ &  $-0.9813$ &  $-1.0037$ &  $-1.0261$ \\
$6$ & $[3/3]$ &  $-0.9125$ &  $-0.9352$ &  $-0.9578$ &  $-0.9803$ &  $-1.0028$ &  $-1.0252$ \\
$6$ & $[4/2]$ &  $-0.9126$ &  $-0.9352$ &  $-0.9578$ &  $-0.9803$ &  $-1.0028$ &  $-1.0253$ \\
$6$ & $[5/1]$ &  $-0.9202$ &  $-0.9425$ &  $-0.9649$ &  $-0.9872$ &  $-1.0096$ &  $-1.0319$ \\
$6$ & $[6/0]$ &  $-0.9416$ &  $-0.9639$ &  $-0.9862$ &  $-1.0085$ &  $-1.0308$ &  $-1.0532$ \\
\hline 
\end{tabular}
\caption{\label{tab:TOTH}
  Pad\'e approximations performed in the variable $z$. No high-energy results
  have been used. Again $n_l$ has been varied from 0 to 5, the dependence,
  however, is very weak.
}
\end{center}
\end{table}
}

Using the results of Tabs.~\ref{tab:TOTL} and~\ref{tab:TOTH} with $n_l=0$
and $n_l$-dependent parts from Eqs.~(\ref{eq:zm3loop})
and~(\ref{eq:resQED3loop})
the formulae~(\ref{eq:zmlog}),~(\ref{eq:zmzm}) and~(\ref{eq:zminv})
can be written in the form
\begin{eqnarray}
  \frac{m(M)}{M} &=&
  1 
  -1.333 
  a_M
  +a_M^2
  \left[
    -14.33 + 1.041 n_l
  \right]
  \nonumber\\&&\mbox{}
  +a_M^3
  \left[
    -202(5) + 27.3(7) n_l  - 0.653 n_l^2
  \right]
  \,,
  \label{eq:zmlog2}
  \\
  \frac{\mu_m}{M} &=&
  1 
  -1.333
  a_M
  +a_M^2
  \left[
    -11.67 + 1.041 n_l
  \right]
  \nonumber\\&&\mbox{}
  +a_M^3
  \left[
    -170(5) + 24.8(7) n_l - 0.653 n_l^2
  \right]
  \,,
  \label{eq:zmzm2}
  \\
  \frac{M}{m(m)} &=&
  1 
  +1.333
  a_m
  +a_m^2
  \left[
    13.44 - 1.041 n_l
  \right]
  \nonumber\\&&\mbox{}
  +a_m^3
  \left[
    194(5) - 27.0(7) n_l + 0.653 n_l^2
  \right]
  \,,
  \label{eq:zminv2}
\end{eqnarray}
with $a_m=\alpha_s^{(n_f)}(m)/\pi$.
For simplicity $\mu=M$ and $\mu=m$ has been chosen
in~(\ref{eq:zmlog2}) and~(\ref{eq:zminv2}), respectively.
In the above equations the exact values of the $n_l^2$ term~\cite{BenBra95}
is displayed. Our method leads to 0.66(2) which is in very good agreement.
This is a further justification for our approach.

For completeness we also want to list the results for the values
$n_l=1,\ldots,5$. They are also obtained from 
Tabs.~\ref{tab:TOTL} and~\ref{tab:TOTH}
and summarized in Tab.~\ref{tab:nl} where  also 
the two-loop coefficients are listed.
Actually the coefficients of the terms linear in $n_l$ 
of Eqs.~(\ref{eq:zmlog2})--(\ref{eq:zminv2}) have been
obtained by performing a fit to the three-loop results of
Tab.~\ref{tab:nl}.  The  errors  of about  2--3\%  
for the three-loop results 
of~(\ref{eq:zmlog2})--(\ref{eq:zminv2}) and Tab.~\ref{tab:nl}
have again been obtained by doubling the spread of Pad\'e approximants.
On one side this is justified with the
behaviour at ${\cal O}(\alpha_s^2)$ where the results of the first
column in Tab.~\ref{tab:nl} are reproduced with the same order of accuracy
(cf. Tab.~\ref{tab:2lfixnl1}).
On the other side the Pad\'e approximants demonstrate more
stability in the case where the moments are added and
$n_l$ is fixed afterwards
than in the case where the Pad\'e procedure is applied to the
individual colour structures separately.
We want to stress that the errors assigned to individual terms in
(\ref{eq:zmlog2})--(\ref{eq:zminv2}) are, in fact, correlated.
Thus for practical applications with fixed values for $n_l$
Tab.~\ref{tab:nl} should be used.

{\footnotesize
  \begin{table}[t]
    \begin{center}
      \begin{tabular}{|l||r|r|r|r|r|r|} 
        \hline
        & \multicolumn{2}{c|}{$m(M)/M$}
        & \multicolumn{2}{c|}{$\mu_m/M$}
        & \multicolumn{2}{c|}{$M/m(m)$}
        \\
        \hline
        $n_l$ 
        & ${\cal O}(\alpha_s^2)$ & ${\cal O}(\alpha_s^3)$
        & ${\cal O}(\alpha_s^2)$ & ${\cal O}(\alpha_s^3)$
        & ${\cal O}(\alpha_s^2)$ & ${\cal O}(\alpha_s^3)$
        \\
        \hline
$0$ &
$    -14.33$ & $   -202(5)$ &
$    -11.67$ & $   -170(5)$ &
$     13.44$ & $    194(5)$ \\
$1$ &
$    -13.29$ & $   -176(4)$ &
$    -10.62$ & $   -146(4)$ &
$     12.40$ & $    168(4)$ \\
$2$ &
$    -12.25$ & $   -150(3)$ &
$     -9.58$ & $   -123(3)$ &
$     11.36$ & $    143(3)$ \\
$3$ &
$    -11.21$ & $   -126(3)$ &
$     -8.54$ & $   -101(3)$ &
$     10.32$ & $    119(3)$ \\
$4$ &
$    -10.17$ & $   -103(2)$ &
$     -7.50$ & $    -81(2)$ &
$      9.28$ & $     96(2)$ \\
$5$ &
$     -9.13$ & $    -82(2)$ &
$     -6.46$ & $    -62(2)$ &
$      8.24$ & $     75(2)$ \\
        \hline
      \end{tabular}
      \caption{\label{tab:nl}
        Dependence of $z_m^{(2)}$ and $z_m^{(3)}$ on $n_l$. The choice
        $\mu^2=M^2$, respectively, $\mu^2=m^2$
        has been adopted.
        }
    \end{center}
  \end{table}
  }

In our calculation we have neglected the effects due to the light
quark masses. At order $\alpha_s^2$ these can be taken into account by
replacing the $n_l$ term in~(\ref{eq:zmlog2}),~(\ref{eq:zmzm2})
and~(\ref{eq:zminv2}) with the following
expression~\cite{GraBroGraSch90}
\begin{equation}
  -1.041 n_l +
  \frac{4}{3}\sum_{1\le i \le n_l}
  \Delta\left(\frac{M_{{i}}}{M}\right)
  \,,
  \label{eq:K:num2}
\end{equation}
where $i$ runs over all light quark flavours.
If $0 \le r \le 1$  then the function
$\Delta(r)$ may be conveniently approximated as follows
\begin{equation}
  \Delta(r) = \frac{\pi^2}{8}~r - 0.597~r^2 + 0.230~r^3
  \,,
  \label{K-appr}
\end{equation}
which is accurate to $1$\%.


\section{Discussion and Applications}
\label{sec:appl}

In this section we compare our results with various
predictions which already exist in the literature. Furthermore a few
applications are discussed where the relation between
the $\overline{\rm MS}$ and on-shell quark mass is explicitly needed.
In particular we will
demonstrate that the new term of ${\cal O}(\alpha_s^3)$ is indeed
necessary to perform consistent analysises in
some practical applications.

In a first step we want to confront the estimations for the ${\cal
  O}(\alpha_s^3)$ terms obtained with the 
help of different optimization procedures with the exact result.
In~\cite{CheKniSir97} the fastest apparent convergence (FAC)~\cite{FAC}
and the principle of minimal sensitivity (PMS)~\cite{PMS}
have been used in order to predict the three-loop coefficient of
$M/m(m)$. In Tab.~\ref{tab:FACPMS} the results are compared with ours.
For $n_l=2$ the discrepancy amounts to only 7\%. It even reduces to only 2\%
for $n_l=5$, i.e. in the case of the top quark.
{\footnotesize
  \begin{table}[t]
    \begin{center}
      \begin{tabular}{|l||r||r|r|r|}
        \hline
        $n_l$ 
        & this work
        & \cite{CheKniSir97} (FAC)
        & \cite{CheKniSir97} (PMS)
        & \cite{BenBra95} (``large-$\beta_0$'')
        \\
        \hline
        $2$ &
        $    143(3)$
        & $152.71$ & $153.76$ & $137.23$\\
        $3$ &
        $    119(3)$
        & $124.10$ & $124.89$ & $118.95$\\
        $4$ &
        $     96(2)$
        & $97.729$ & $98.259$ & $101.98$\\
        $5$ &
        $     75(2)$
        & $73.616$ & $73.903$ & $86.318$\\
        \hline
      \end{tabular}
      \caption{\label{tab:FACPMS}Comparison
        of the results obtained in this
        paper with estimates based of FAC, PMS and the ``large-$\beta_0$''
        approximation for $M/m(m)$.
        }
    \end{center}
  \end{table}
  }

In Eq.~(\ref{eq:zminv2}) it is also possible to use the quantity 
$\alpha_s(M)$ as an expansion parameter (instead of $\alpha_s(m)$).
For $n_l=5$ the three-loop coefficient then reads $81.45(72)$ whereas the
coefficient at ${\cal O}(\alpha_s^2)$ does not change.
For this case the authors of~\cite{CheKniSir97} have obtained $79.838$ which
deviates by roughly 2\%. This coefficient has also been estimated
in~\cite{Sir95} to be $76.17$ which agrees with our central value 
within approximately 6\%.

Let us compare our results also with the ones obtained in the large
$\beta_0$-limit~\cite{BenBra95},
where $\beta_0$ is the first coefficient of the QCD $\beta$ function.
The results obtained in~\cite{BenBra95} can also be found in
Tab.~\ref{tab:FACPMS}. 
Excellent agreement below 1\% is found for $n_l=3$. It amounts to
roughly 5\% for $n_l=4$ and 14\% for $n_l=5$.

In the remaining part of this section we want to discuss the application of
our results in the context of quark production close to the threshold.
Calculations in this context are connected to the pole mass of the involved
quarks. 
To be specific let us consider the production of top quarks in
$e^+e^-$ collisions.
The corresponding physical observables
expressed in terms of $M_t$ show in general a bad
convergence behaviour. In the case of the total cross section, e.g., the
next-to-next-to-leading order corrections partly exceed the next-to-leading
ones. Furthermore the peak position which is the most striking feature of the
total cross section and from which finally the mass value
can be extracted depends very much on the number of terms one includes into
the analysis.
The commonly accepted explanation for this is that
the pole mass is sensitive to long-distance effects which result to
intrinsic uncertainties of order $\Lambda_{QCD}$~\cite{BenBra94,Big94}.
In other words, it is not possible to determine the pole mass from the
analysis of the cross section at threshold with an accuracy better than
$\Lambda_{QCD}$. 

Several strategies have been proposed to circumvent this
problem~\cite{Ben98,HoaSmiSteWil98,HoaTeu99}.
They are based on the observation that
the same kind of ambiguities also appear in the static quark
potential, $V(r)$. In the combination $2M_t + V(r)$, however, the
infra-red sensitivity drops out. Thus a definition of a short-distance
mass extracted from threshold quantities should be possible.
The relation of the new mass parameter
to the pole mass is used in order to re-parameterize the threshold phenomena.
On the other hand a relation of the new quark mass to the $\overline{\rm MS}$
mass must be established as
it is commonly used for the parameterization of those quantities which are not
related to the threshold.
In order to do this consistently
the three-loop relation between the
$\overline{\rm MS}$ and the on-shell mass is needed.

In~\cite{Ben98} the concept of the so-called potential mass, $m_{t,PS}$, has
been introduced. Its connection to the pole mass is given by
\begin{eqnarray}
  m_{t,PS}(\mu_f) &=& M_t - \delta m_t(\mu_f)
  \,,
  \label{eq:mPSdef}
\end{eqnarray}
where $\delta m_t(\mu_f)$ is connected to the potential in momentum space,
$\tilde{V}(q)$, through the equation
\begin{eqnarray}
  \delta m_t(\mu_f) &=& -\frac{1}{2} \int_{|\vec{q}|<\mu_f}
  \frac{{\rm d} \vec{q}}{\left(2\pi\right)^3}\tilde{V}(\vec{q})
  \,.
\end{eqnarray}
Via this equation a subtracted potential, $V(r,\mu_f)$, is defined.
The factorization scale $\mu_f$ has been introduced in order to extract
the infra-red behaviour arising from the potential.
In the combination~(\ref{eq:mPSdef})
it cancels against the one of $M_t$ leading to an enormous
reduction of the long-distance uncertainties in $m_{t,PS}$~\cite{Ben98}.
Thus it is promising to formulate the threshold problems in terms of
$m_{t,PS}(\mu_f)$ and $V(r,\mu_f)$ instead of $M_t$ and $V(r)$.
In the final result the dependence on $\mu_f$ cancels.
For the numerical analysis the value $\mu_f=20$~GeV has been adopted
in~\cite{BenSigSmi99} as its upper bound is roughly given by
$M_tC_F\alpha_s(\mu)$.

The computation of $\delta m_t(\mu_f)$ can be performed up to ${\cal
  O}(\alpha_s^3)$ using the explicit results for
$\tilde{V}(\vec{q})$~\cite{Sch98}. The result reads
\begin{eqnarray}
  \delta m_t(\mu_f) &=& C_F\mu_f\frac{\alpha_s}{\pi}\left\{
    1
    +\frac{\alpha_s}{\pi}\left[
      \frac{97}{12} - \frac{11}{4}\logfu
      + n_l\left(-\frac{11}{18} + \frac{1}{6}\logfu\right) 
    \right]
    +\left(\frac{\alpha_s}{\pi}\right)^2\left[
      \frac{33623}{288} 
      + \frac{9}{4}\zeta_2
      \right.\right.\nonumber\\&&\left.\left.\mbox{}
      + \frac{33}{8}\zeta_3
      - \frac{9}{64}\zeta_4
      - \frac{305}{6}\logfu 
      + \frac{121}{16}\logfu^2 
      + n_l\left(
        -\frac{7145}{432} 
        - \frac{13}{12}\zeta_3
        + \frac{493}{72}\logfu
      \right.\right.\right.\nonumber\\&&\left.\left.\left.\mbox{}
        - \frac{11}{12}\logfu^2
        \right) 
      + n_l^2\left(
        \frac{157}{324} - \frac{11}{54}\logfu + \frac{1}{36}\logfu^2
        \right)
  \right]
  \right\}
  \label{eq:deltam}
  \,,
\end{eqnarray}
with $\alpha_s=\alpha_s^{(6)}(\mu)$ and $\logfu=\ln\mu_f^2/\mu^2$.
We are now in the position to establish a relation between
the two short-distance masses $m_{t,PS}(\mu_f)$ and $m_t(\mu)$.
Inserting Eqs.~(\ref{eq:deltam}) and~(\ref{eq:zminv2})
into~(\ref{eq:mPSdef}) we get:
\begin{eqnarray}
  m_{t,PS}(20~{\rm GeV}) &=& \left( 165.0 + 6.7 + 1.2 + 0.28
  \right)~\mbox{GeV}
  \,,
  \label{eq:mtPSmtmt}
\end{eqnarray}
where the different terms represent the contributions form order $\alpha_s^0$
to $\alpha_s^3$. For the numerical values
$m_t(m_t)=165.0$~GeV and $\alpha_s^{(6)}(m_t(m_t))=0.1085$
have been used.
Note that the error of the ${\cal O}(\alpha_s^3)$ coefficient
in the $\overline{\rm MS}$--on-shell mass relation is negligible.
The comparison of Eq.~(\ref{eq:mtPSmtmt}) with the analogous expansion
for $M_t$,
\begin{eqnarray}
  M_t &=& ( 165.0 + 7.6 + 1.6 + 0.51 )\mbox{~GeV}
  \,,
\end{eqnarray}
shows that the potential mass can be 
more accurately related to the $\overline{\rm MS}$ mass than $M_t$.

The last term of the expansion in~(\ref{eq:mtPSmtmt})
is of the same order of magnitude as the error in the top quark mass
determination at a next-linear-collider.
Whereas in~\cite{BenSigSmi99} this term 
has been taken as uncertainty in the mass relation
the error reduces significantly after the knowledge of the
${\cal O}(\alpha_s^3)$ term of the $\overline{\rm MS}$--on-shell relation.
This can be deduced from the well-behaved expansion in
Eq.~(\ref{eq:mtPSmtmt}).
The dominant error is now provided by the uncertainty in $\alpha_s$.

A similar strategy has been proposed in~\cite{HoaTeu99}.
There the so-called 1S~mass, $M_t^{1S}$, was defined as half the perturbative
mass of a fictious toponium $1^3S_1$ ground state
which would exist if the top quark were stable.
The relation between $M_t^{1S}$ and $M_t$ is given by
\begin{eqnarray}
  M_t^{1S} &=& M_t - \frac{M_t C_F^2 \alpha_s^2}{8} \left\{
    \epsilon
    + \epsilon^2 \frac{\alpha_s}{\pi} \left[
      \frac{97}{6} + 11\lmucft 
      + n_l\left(-\frac{11}{9} - \frac{2}{3}\lmucft\right)
    \right]
    + \epsilon^3 \left(\frac{\alpha_s}{\pi}\right)^2 \left[
      \frac{1793}{12} 
      \right.\right.\nonumber\\&&\left.\left.\mbox{}
      + \frac{2917}{216}\zeta_2 
      + \frac{275}{4}\zeta_3
      - \frac{9}{32}\zeta_4
      + \frac{927}{4}\lmucft 
      + \frac{363}{4}\lmucft^2 
      + n_l\left(
        - \frac{1693}{72}
        - \frac{11}{18}\zeta_2
        - \frac{19}{2}\zeta_3
      \right.\right.\right.\nonumber\\&&\left.\left.\left.\mbox{}
        - \frac{193}{6}\lmucft
        - 11\lmucft^2 
      \right)
      + n_l^2\left(
        \frac{77}{108}
        + \frac{1}{54}\zeta_2
        + \frac{2}{9}\zeta_3
        + \lmucft 
        + \frac{1}{3} \lmucft^2 
      \right)
    \right]
  \right\}\,,
  \label{eq:M1SMt}
\end{eqnarray}
with $\alpha_s=\alpha_s^{(6)}(\mu)$ and $\lmucft=\ln(\mu/(C_F\alpha_s M_t))$.
The philosophy is very similar as in the case of the potential mass.
From the experiment the quantity $M_t^{1S}$ is extracted.
In~\cite{HoaTeu99} it has been shown that this is possible with an uncertainty
of approximately $200$~MeV. In a next step $M_t^{1S}$ has to be related to
the $\overline{\rm MS}$ mass $m_t(m_t)$.
As the extraction of $M_t^{1S}$ is based on a
next-to-next-to-leading order formalism the ${\cal O}(\alpha_s^3)$ relation
computed in this work is necessary.

In practice one proceeds as follows: In a first step~(\ref{eq:M1SMt})
is inverted and then equated to~(\ref{eq:zminv2}).
After setting $\mu=m_t(\mu)$ it is possible to solve the equation numerically
for $m_t(m_t)$.
Care has to be taken in connection to the expansion parameter in
Eq.~(\ref{eq:M1SMt}). A naive counting in powers of $\alpha_s$ would lead to
theoretically inconsistent results as was demonstrated in~\cite{HoaLigMan99}.
A consistent treatment is obtained if an expansion parameter $\epsilon$ is
introduced where the terms of order $\alpha_s^n$ are
proportional to $\epsilon^{n-1}$ in Eq.~(\ref{eq:M1SMt}) and 
proportional to $\epsilon^{n}$ in Eq.~(\ref{eq:zminv2}).

One finally arrives at the following relation between the
$\overline{\rm MS}$ and $1S$~mass
\begin{eqnarray}
  m_t(m_t) &=& \left( 175.00 - 7.60 - 0.97 - 0.14 \right)~{\rm GeV}
  \label{eq:mtmtM1S}
  \,.
\end{eqnarray}
where $M_t^{1S}=175$~GeV and $\alpha_s^{(5)}(M_Z)=0.118$ has been adopted.
Using the large-$\beta_0$ results for the order $\alpha_s^3$
term of~(\ref{eq:zminv2}) the last term reads $-0.23$~\cite{HoaTeu99}
which is off by more than 50\% form the exact result.
The conclusions which can be drawn from Eq.~(\ref{eq:mtmtM1S}) are very
similar to the ones stated above: the uncertainties due to unknown terms in
the mass relations are negligible as compared to the error with which
$M_t^{1S}$ can be extracted from the experiment. The dominant uncertainty
comes from the error in $\alpha_s$ which amounts for $\pm0.003$ to roughly
200~MeV~\cite{HoaTeu99} in Eq.~(\ref{eq:mtmtM1S}).

For completeness we would like to mention that the introduction of short
distance masses like $m_{t,PS}$ or $M_t^{1S}$ have significantly 
stabilized the position of the peak in the top quark cross section
which is important in extracting the mass value.
The normalization uncertainty, however, remains large when including higher
order corrections.

Also the bottom quark mass can be extracted from
quantities related to the quark threshold.
Recently~\cite{PenPiv98,MelYel99PenPiv99BenSig99,Hoa99}
a precise value for the bottom quark mass has
been determined in the context of QCD sum rules.
For example, in~\cite{Hoa99} the on-shell mass was eliminated in
favour of the 1S mass in order to reduce the error.
Once $M_b^{1S}$ is determined the analogous equation to~(\ref{eq:M1SMt})
in combination with~(\ref{eq:zminv2}) can be used in order to get
$m_b(m_b)$.
In~\cite{Hoa99} the values $M_b^{1S}=4.71\pm0.03$~GeV and 
$m_b(m_b)=4.2\pm0.06$ have been obtained where the large-$\beta_0$
approximation for the three-loop term in the $\overline{\rm
MS}$--on-shell relation is part of the error in $m_b(m_b)$.
Following the procedure described in~\cite{Hoa99} one arrives at
\begin{eqnarray}
  m_b(m_b) &=& \left( 4.71 - 0.40 - 0.11 - 0.03 \pm 0.03 \pm 0.04
               \right)~\mbox{GeV}       
  \label{eq:mbmbM1S}
  \,,
\end{eqnarray}
where the different terms correspond to different orders in the
$\Upsilon$-expansion.
The first error is due to $M_b^{1S}$ and the second one reflects
the error in $\alpha_s(M_Z)=0.118\pm0.004$ which is adopted
from~\cite{Hoa99}. Due to the nice convergent behaviour
of~(\ref{eq:mbmbM1S}) the total error on $m_b(m_b)$ only contains these
two sources which finally leads to
\begin{eqnarray}
  m_b(m_b) = 4.17 \pm 0.05~\mbox{GeV}
  \,.
  \label{mb(mb)_from_M1S:final}
\end{eqnarray}
It is important to stress that taking into account of the newly
computed ${\cal O}(\alpha_s^3)$ term in the $\overline{\rm
MS}$--on-shell relation is crucial for the reliable estimation of
the errors in (\ref{mb(mb)_from_M1S:final}). Indeed,  a deviation of 
the real value for $z_m^{(3)}$ from the large-$\beta_0$ estimation 
by, say, a factor of two, which one could not exclude a priori,
would result to a systematic shift in  $m_b(m_b)$
of around  100~MeV.

A somewhat different approach for the determination of both the charm and
bottom mass has been followed in~\cite{PinYnd98}.
There the  lower states in the heavy quarkonium spectrum were  computed
up to order $\alpha_s^4$.
This allows the extraction of relatively accurate values for
the pole masses of the charm and bottom quarks.
The transformation to the $\overline{\rm MS}$ mass has been performed with the
help of the two-loop relation~\cite{GraBroGraSch90}.
However, to the order the quarkonium spectrum was computed
it is more consistent to use the ${\cal O}(\alpha_s^3)$ relation provided in
this paper.

{\footnotesize
  \begin{table}[t]
    \begin{center}
      \begin{tabular}{|l||r|r|r|}
        \hline
        accuracy
        & ${\cal O}(\alpha_s^2)$
        & ${\cal O}(\alpha_s^3)$
        & ${\cal O}(\alpha_s^4)$
        \\
        \hline
        $M_b$ (GeV) 
        & 4.752 & 4.858 & 5.001 \\
        $m_b(m_b)$ (GeV) 
        & 4.362 & 4.292 & 4.322 \\
        \hline
      \end{tabular}
      \caption{\label{tab:Mbmb}Bottom quark mass determined with increasing
        accuracy.
        }
    \end{center}
  \end{table}
  }

In fact, in~\cite{PinYnd98} the pole mass of the bottom quark is given with a
accuracy of order $\alpha_s^2$, $\alpha_s^3$ and $\alpha_s^4$.
For the conversion to the $\overline{\rm MS}$ scheme relation~(\ref{eq:zmzm2})
is needed up to one, two and three loops, respectively.
The same accuracy is required for the running of $\alpha_s$ from $M_Z$ to
$M_b$. Our results are displayed in Tab.~\ref{tab:Mbmb}. As in~\cite{PinYnd98}
the value $\alpha_s^{(5)}(M_Z)=0.114$ has been adopted.
The numerical values slightly differ from the ones given in~\cite{PinYnd98}
which can be traced back to a minor inconsistent treatments
in~\cite{PinYnd98}. Thus taking over the error estimates from~\cite{PinYnd98}
the on-shell value for the bottom quark mass reads
$M_b = 5.001^{+0.104}_{-0.066}\mbox{~GeV}$.
It transforms to the following value for $m_b(m_b)$
\begin{eqnarray}
  m_b(m_b) &=& 4.322^{+0.043}_{-0.028}\mbox{~GeV}
  \,.
\end{eqnarray}

In a similar way also the results for the charm quark mass can be obtained.
In addition to the bottom quark case the matching between four and five
flavours has to be performed consistently. This means that $n$-loop running
has to be accompanied with $(n-1)$-loop matching.
For a detailed description we refer to~\cite{CheKniSte97}.
The on-shell value~\cite{PinYnd98}
\begin{eqnarray}
  M_c &=& 1.866^{+0.190}_{-0.154}\mbox{~GeV}
  \,,
\end{eqnarray}
leads to 
\begin{eqnarray}
  m_c(m_c) &=& 1.377^{+0.132}_{-0.127}\mbox{~GeV}
  \,,
\end{eqnarray}
where again the error estimates are taken over from~\cite{PinYnd98}.

Compared to~\cite{PinYnd98} the inclusion of the
${\cal O}(\alpha_s^3)$ terms leads to a shift in the central
values of more than $100$~MeV.
In the case of the bottom quark this change is even larger than
the errors presented in~\cite{PinYnd98} which
might indicate that the estimates were too optimistic.
This demonstrates that a consistent treatment of the different orders in
$\alpha_s$ is absolutely crucial.

We would like to mention that
in this paper we don't intend to compare and judge the different methods used
for extracting the bottom quark mass and the corresponding error estimate.
For a comprehensive discussion of recent determinations of
$m_b(m_b)$ we refer to the third reference
in~\cite{MelYel99PenPiv99BenSig99}.


\section{Conclusions}
\label{sec:con}
In this paper the three-loop term of order $\alpha_s^3$ to the relation
between the $\overline{\rm MS}$ and on-shell definition of the quark mass
is computed. This is achieved by considering expansions of the quark self
energy for small and large external momentum followed by a conformal
mapping and a Pad\'e approximation.

The three-loop computation in the limit of small momentum requires
the knowledge of integrals which before have never been used in practical
applications. We have checked the results available in the
literature and added analytical results for the missing integrals.

The procedure used for the computation 
is tested in detail at one- and two-loop order and compared to the known
result. This also provides an estimate on the uncertainty
which is between 2~and 3\% for the order $\alpha_s^3$ contribution.

We discuss several important applications of the newly available term.
In particular the relations between the $\overline{\rm MS}$ mass and
other short-distance masses are considered. It is shown that the
three-loop term is necessary to reduce the error with which finally the 
$\overline{\rm MS}$ quark mass can be determined.

In conclusion we would also like to stress that our procedure is not
limited to the determination of the conversion factor between the
$\overline{\rm MS}$ and on-shell quark masses. As a spin-off we also
got a wealth of information about the (perturbative) quark propagator
in QCD at order $\alpha_s^3$. It should be possible to use this
approach in order to find an accurate semianalytic description of the
propagator in  all possible regions of momentum transfer and for 
a general gauge fixing condition.


\section*{Acknowledgments}

We would like to thank P.A.~Grassi and A.H.~Hoang for useful conversation
and J.H.~K\"uhn and K.~Melnikov for
valuable suggestions.
One of authors (K.G.Ch.) thanks A.A.~Pivovarov for critical remarks.
This work was supported by DFG under Contract Ku 502/8-1
({\it DFG-Forschergruppe ``Quantenfeldtheorie, Computeralgebra und
  Monte-Carlo-Simulationen''}).


\section*{Note added}

An  analytical reevaluation of the relation between
$\overline{\rm MS}$ and on-shell quark masses have been reported
recently in \cite{MelRit99}. Their result  is in full
agreemeent 
with our Eqs.~(\ref{eq:zmlog2})--(\ref{eq:zminv2}) and our 
Tabs.~\ref{tab:TOTL},~\ref{tab:TOTH},~\ref{tab:nl} and ~\ref{tab:FACPMS}.
It reads:
\begin{eqnarray}
  \frac{m(M)}{M} &=& 1 
  - \frac{4}{3}\frac{\alpha_s}{\pi}
  + \left(\frac{\alpha_s}{\pi}\right)^2 \left(
    -14.3323 + 1.0414 n_l
  \right)
  \nonumber\\&&\mbox{}
  + \left(\frac{\alpha_s}{\pi}\right)^3 \left(
    -198.7068 + 26.9239 n_l - 0.65269 n_l^2
  \right)
  \nonumber
  \,.  
\end{eqnarray} 
The results of~\cite{MelRit99} for all ten independent colour
structures appearing in the $\alpha_s^3$ order are also in full
agreement\footnote{after correcting a few misprints in the final
equations of~\cite{MelRit99}} with ours with the only exception of the
colour structure $FAA$. For this case our estimation of the error bars
($z_m^{FAA}=-16.4(3)$ versus $z_m^{FAA}=-15.85$ as found
in~\cite{MelRit99}) seems to  be sligtly too optimistic (provided, of
course, that the very calculation of~\cite{MelRit99} gives the correct
answer for this colour structure).


\end{document}